\documentclass[preprint]{aastex}

\shortauthors{Gizis et al.}
\shorttitle{New Neighbors}

\begin{document}

\title{New Neighbors from 2MASS: Activity and Kinematics at the 
Bottom of the Main Sequence}

\author{John E. Gizis\altaffilmark{1}}
\affil{Infrared Processing and Analysis Center, 100-22, 
  California Institute of Technology,
  Pasadena, CA 91125\email{gizis@ipac.caltech.edu}}
\author{David G. Monet\altaffilmark{1}}
\affil{U.S. Naval Observatory, P.O. Box 1149, Flagstaff, AZ 86002}
\author{I. Neill Reid}
\affil{Department of  Physics and Astronomy, University of Pennsylvania,
209 South 33rd Street, Philadelphia PA 19104-6396}
\author{J. Davy Kirkpatrick}
\affil{Infrared Processing and Analysis Center, 100-22, 
  California Institute of Technology,
  Pasadena, CA 91125 }
\author{James Liebert}
\affil{Steward Observatory, University of Arizona, Tucson AZ 85721}
\author{Rik J. Williams}
\affil{Department of Astronomy, MSC 152, California Institute of Technology, 
Pasadena, CA 91126-0152}

\altaffiltext{1}{Visiting Astronomer, Kitt Peak National Observatory, 
National Optical Astronomy Observatories, which is operated by
the Association of Universities for Research in Astronomy, Inc. 
(AURA) under cooperative agreement with the National Science Foundation.}

\begin{abstract}
\footnotesize
We have combined 2MASS and POSS II data in a search for 
nearby ultracool (later than M6.5) 
dwarfs with $K_s<12$.  Spectroscopic follow-up observations
identify 53 M7 to M9.5 dwarfs and seven L dwarfs.  The observed space
density is $0.0045 \pm 0.0008$ M8-M9.5 dwarfs per cubic parsec,  
without accounting for biases, consistent
with a mass function that is smooth across the stellar/substellar
limit.  
We show the observed frequency of H$\alpha$ emission peaks at
$\sim 100\%$ for M7 dwarfs and then decreases for cooler dwarfs.  
In absolute terms, however, as measured by the ratio of 
H$\alpha$ to bolometric luminosity,
none of the ultracool M dwarfs can be considered very active
compared to earlier M dwarfs, and we show that the decrease that
begins at spectral type M6 continues to the latest L dwarfs.  
We find that flaring is common among the coolest M dwarfs and
estimate the frequency of flares at 7\% or higher.  
We show that the kinematics of relatively active 
($EW_{H\alpha}>6$ \AA) ultracool M dwarfs are
consistent with an ordinary old disk stellar population, while
the kinematics of inactive ultracool M dwarfs are more typical of a
0.5 Gyr old population.  
The early L dwarfs in the sample have kinematics consistent
with old ages, suggesting that the hydrogen burning limit
is near spectral types L2-L4.  
We use the available data on M and L dwarfs to show that 
chromospheric activity drops with decreasing mass and temperature,
and that at a given (M8 or later) spectral type, the 
younger field (brown) dwarfs are
less active than many of the older, more massive field stellar dwarfs.
Thus, contrary to the well-known stellar age-activity relationship, 
low activity in field ultracool dwarfs can be an indication of
comparative youth and substellar mass.
\end{abstract}


\section{Introduction \label{intro}}

Catalogs of nearby stars \citep{gj91,khs95,rhg95} 
and high proper motion stars \citep{l79} 
are grossly deficient in very low mass (VLM) dwarfs.  
With spectral types of M7 and later,
these objects, sometimes called ``ultracool M dwarfs,''\footnote{
All spectral types in this paper are on the \citet{khm91} 
M dwarf and \citet{k99} L dwarf systems.  L dwarfs are
cooler than ``ultracool'' M dwarfs.} are so optically faint
that even nearby ones eluded searches based on the older (pre-1980s) 
sky surveys.
These dwarfs have particular importance because they lie at or
below the hydrogen burning limit -- and have proven not
only to be estimators of the numbers of dark brown dwarfs,
but also present interesting astrophysical challenges in
their own right.  

The new generation of sky surveys allows this deficiency to be addressed
and large samples of nearby VLM dwarfs to be identified.
The Two Micron All-Sky Survey\footnote{2MASS data and 
documentation are available at 
\url{http://www.ipac.caltech.edu/2mass}}
 (Skrutskie et al., in prep.; hereafter 2MASS) 
provides reliable photometry in the JHK$_s$ passbands, close to
the peak of emission for these cool dwarfs.  Furthermore, the Second
Palomar Sky Survey \citep{poss2}, hereafter POSS II) provides 
B$_J$,R$_{F}$, and I$_N$ photographic
photometry in the northern hemisphere.  In the southern hemisphere, the
UK Schmidt and ESO sky survey plates provide $B_J$ and $R_F$ magnitudes.  
In sum, it is becoming possible to identify both the least luminous
stars and young massive brown dwarfs by their optical and 
near-infrared colors alone over most of the sky.  

We present first results of a search using near-infrared and optical
sky survey data aimed at completing the nearby star catalog 
for the ultracool M dwarfs.
We discuss the sample selection and spectroscopic followup in
Section~\ref{data}.   Although the sample discussed in this paper 
includes only a small fraction of the total population of nearby ultracool 
dwarfs,  it represents a fourfold increase in the number of such sources 
known. We discuss some preliminary results concerning the 
statistical properties of these
sources in the latter sections of this paper.  We discuss 
some stars of special interest in Section~\ref{special}
and the 2MASS colors of ultracool M dwarfs in Section~\ref{colors}.  
The local space density of VLM dwarfs
is discussed in Section~\ref{lf}, their 
activity and kinematics are
discussed in Section~\ref{activity}, and finally 
our conclusions and future prospects are discussed in ~\ref{summary}.

\section{Data \label{data}}

\subsection{Sample Selection \label{sample}}

Our results are based on three observational samples.  For 
our initial observing run, in July 1998,
we used both photometric and proper motion 
criteria to define a sample of candidate VLM dwarfs. Based on 
the results from this run and further
experience analyzing 2MASS data, we were able to improve 
our selection criteria for our December 1998 and subsequent
observing runs.  There are thus three samples with different
properties, and when necessary we distinguish them as
``Sample A'' (July 1998), ``Sample B'' (December 1998), 
and  ``Sample C'' (June 1999) respectively.  Samples B and C
have nearly identical selection criteria, and when combined
are refered to as Sample BC.
In all samples, objects within 20 degrees of the Galactic
plane were excluded.  Unless otherwise stated, all the analysis
in Section~\ref{lf} and~\ref{activity} is based on Sample BC.  

Sample A was based on 2MASS data processed by July 1998.  
All these data were obtained at the Mt. Hopkins 2MASS telescope
and a total of 363 square degrees
were searched.   All objects classified
as extended by the 2MASS pipeline were eliminated (\cite{jarrett}).  
The 2MASS data were correlated 
with the USNO's PMM scans of the POSS II plates, using a preliminary
version of the software used by the 2MASS Rare Objects Core
Project (this software will be described in more detail 
in a future publication by Monet et al.).  This provided
three additional colors: B$_J$, R$_F$, and I$_N$.  Zero-point
calibrations for the POSS II scans were not available, but
we have since found that rough zero-points (good to $\pm 0.5$ magnitudes)
are $B = B_{J inst} +2$, $R_C = R_{F inst} -1$, and 
$I_C = I_{N inst} -1$.  These do not account for 
plate-to-plate variations or the significant color terms 
expected in $B_J$ and $R_F$.  

All objects that met the following criteria were observed:

\begin{enumerate}
\item $9.0 \le$ K$_s \le 13.0$
\item $0.95 \le$ J-K$_s < 1.30 $
\item no B$_J$ detection
\item Significant ($>2 \sigma$) proper motion 
\end{enumerate}

Objects were also observed if they satisfied:
\begin{enumerate}
\item $9 \le $ K$_s \le 12.0$
\item $0.95 \le$ J-K$_s < 1.30 $
\item B$_J$ detection
\item Significant ($>2 \sigma$) proper motion
\end{enumerate}

The proper motion criterion requires elaboration.  The 
magnitude of the observed positional offset between the 2MASS and POSS II
source is compared to the distribution of all 2MASS-POSS II correlations.
Only objects with significant proper motion were selected.  
Thus, while we selected sources that are moving with high confidence
(roughly $2 \sigma$), the actual cutoff in terms of arcseconds per year depends
on the epoch difference between the F plate and 2MASS, which varies
between 0 and $\sim 10$ years.

As shown in Section~\ref{spectra}, these criteria led to the identification
of a number of new nearby M dwarfs, but they are flawed in some respects.  
First, the POSS II and UKST IIIa J plates are sufficiently sensitive that 
many nearby (bright) VLM dwarfs are detected.   Moreover,
the blue cutoff in J-K$_s$ allowed mid-M dwarfs to enter the
sample.  We selected this blue cutoff because
the prototypical M7 dwarf VB 8 has J-K$=0.95$ \citep{l92}.
However, M dwarfs between spectral type M0 and M6.5 all
have J-K$_s \approx 0.9$,
and, with uncertainties of 
$\sigma_{J-K} \approx 0.04$
mag, significant numbers are scattered to  
J-K$_s > 0.95$.  Consequently, our initial attempts to select 
ultracool M dwarfs produced a sample which is 
heavily contaminated by distant mid-M dwarfs.  As it turned out,
there were no 2MASS sources meeting our color and magnitude 
cuts that were not paired with a POSS II source in Sample A.

Based on experience gained from this analysis, we revised our sample 
selection for the December 1998 observing run.   
Because a larger area was available, we focused on 
brighter VLM dwarfs which should lie within $\sim 20$ parsecs.
The Sample B criteria are:

\begin{enumerate}
\item K$_s \le 12.0$
\item J-K$_s \ge 1.00$
\item R$_F$ - K$_s > 3.5$ or I$_N$ - K$_s \ge 2.0$ 
\item $\delta < +30\deg$
\item $\alpha<13^h00^m$ or $\alpha>20^h00^m$ 
\item J-H $\le \frac{4}{3}$ H-K $ + 0.25$
\end{enumerate}

No selection based on proper motion was applied, and therefore
Sample B is kinematically unbiased.  A total of
2977 sq. degrees were searched.  The J-K$_s$
cutoff excludes early and mid-M dwarfs, but also some of the
bluer M7 dwarfs.  Nearby bright L dwarfs, however, are included 
in these samples, since we impose no red cutoff. 
Both samples are magnitude selected, and therefore are biased towards
overluminous stars and unresolved near-equal luminosity binaries, although  
binaries with separations of a few arcseconds may be excluded by the 
extended source provision.  The position selection is due to the 
requirement that the objects be observable from Las Campanas in December.
Ultracool M dwarfs have R-K$_s > 5.5$ and I-K$>3.4$ \citep{l92}, but we used
a more liberal selection to allow for uncertainties in the 
calibration of the photographic magnitudes. 
In Section~\ref{colors}, we show that R-K$_s > 4.9$ includes
all M8 and later dwarfs.  The J-H,H-K cut excludes M giants.  

Sample C was selected for our June 1999 Kitt Peak observing run.
The selection was identical to the Sample B selection, except
that a different area was covered and only objects with $R-K_s > 4.9$
were selected.  Processed data which lay within the following
limits were selected:
\begin{enumerate}
\item $\alpha>11^h00^m$ and $\delta>+6\arcdeg$
\item $16^h20^m< \alpha <23^h35^m$ and $-36 < \delta<+6\arcdeg$
\end{enumerate}

\subsection{Spectroscopy and Data Analysis\label{spectra}}

Sample A was observed on UT dates July 30 -- August 1 1998 
using the Double Spectrograph and the Hale 200-in. telescope
during an observing run that was primarily devoted to 
our ongoing spectroscopic survey of Luyten high proper motion 
stars \citep{gr97}.  
The wavelength coverage included 6290 to 8800 \AA~ at a resolution of
3 \AA.  

Sample B was observed on December 2 -- 7 1998
using the Modular Spectrograph on the Las Campanas 100-in
telescope.  The ``Tek 5'' chip, a 2048-square CCD with $24\mu$ 
pixels, was used with a 600 l/mm
grating blazed at 7500 \AA.  The useful wavelength range of the
spectra was 6100 - 9400 \AA~ at a resolution of 6 \AA.  
A few targets (including all
three objects with J-K$_s>1.3$, which had been previously identified
as L dwarf candidates) were observed 
during Keck observing runs 
(see Kirkpatrick et al. 1999b, hereafter K99, 
and Kirkpatrick et al. 2000, hereafter K00) using LRIS \citep{lris}.
The resolution was 9 \AA~with wavelength
coverage from 6300 to 10100 \AA.  Observations of the flare dwarf
2MASSI J0149089+295613 have already been described in \citet{superflare}.  
The H$\alpha$ activity levels adopted here are the average quiescent
values.  The known object BRI 1222-1222 was not observed,
and we rely on the spectroscopic observations reported
by \citet{khs95} and \citet[hereafter TR]{tr98}.
2MASSW J0354013+231633 has been previously published as 2MASP J0354012+231635
\citep{kbs97}
but the spectral observations presented here are independent.

Sample C was observed June 22 -- 23 1999 using the R.C. spectrograph and the 
Kitt Peak 4m telescope.  The wavelength coverage was 6140 \AA~to 9200 \AA~
using the 2048 CCD, but the extreme ends of the spectra 
were out of focus.  A few objects were observed at Palomar 200-in.
in May 1999.  The four new L dwarfs identified with the Kitt Peak
data were reobserved at higher signal-to-noise using Keck in 
July 1998.  The spectral measurements used for classification on the
K99 system are given in Table~\ref{table-ldwarfs}.  

All spectra were extracted and flux calibrated using IRAF.  
M dwarf spectral types were measured by overplotting 
dwarfs of known spectral type, and should be good to 
$\pm 0.5$ subclasses.  All L dwarfs have Keck observations 
and were classified as in K99 and K00.  
A few M dwarfs in Table~\ref{table-data}
have classifications that differ by 0.5 subclasses from previously
published values -- we leave our values unalterred as 
representative of our uncertainties.  
H$\alpha$ fluxes were measured assuming the data were
photometric (there were no clouds for our observations in
December 1998).
We assume that BC$_K = 3.2$ as derived by \citet{tmr93}
for ultracool M dwarfs.    
We believe the H$\alpha$ fluxes should be viewed with caution
since slit losses and the high airmass of many of the observations
will increase the uncertainties (the spectrograph was not adjusted
to the parallactic angle).  However, our derived
H$\alpha$ to bolometric luminosity ratios are consistent with
those of TR, and in any case the H$\alpha$ 
emission strength is variable in these dwarfs.  
In Table~\ref{table-fluxes}, we list our measured and
derived parameters for the ultracool dwarfs in Sample BC.

Proper motions were estimated by measuring positions off the DSS
(POSS I/UKST J) and XDSS (POSS II/UKST R) images of the
 photographic sky surveys.
When a second epoch photographic sky survey was not available,
we used the 2MASS images instead.  
All motions reported are relative to other stars in the field,
but the correction to absolute motions is negligible compared
to other sources of error in the kinematics.  
The proper motion reported for
LHS2397a is from \citet{l79} and BRI1222-1222 is from \citet{t96}.  
Note that all of our targets are visible on the
DSS images, but most were previously unrecognized.  
The targets which had no POSS II pairings in our initial processing
all proved to have high proper motions.  They are visible on the XDSS
but lie outside the 8 arcsecond search radius employed in cross-referencing
against the photographic data.
Since the 2MASS positions are highly accurate, and
both the 2MASS images and DSS images are (or will shortly be) easily accessible
electronically, we are not presenting finding charts.  

\begin{figure}
\epsscale{0.8}
\plotone{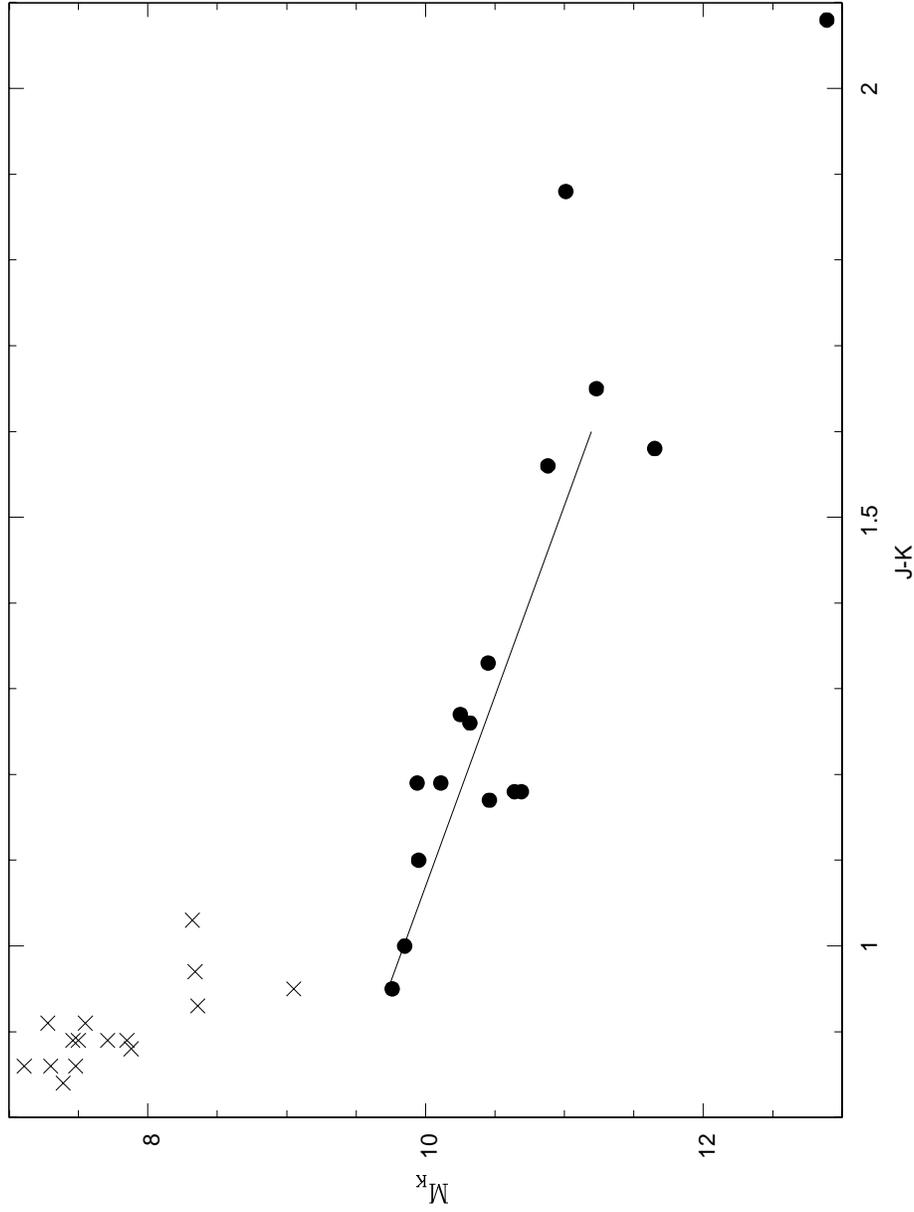}
\caption{Absolute magnitudes for M7 and later dwarfs (solid points) 
with the linear fit 
$M_K = 7.593 + 2.25\times (J-K_s)$ shown.  The scatter about this fit is
$\sigma = 0.36$ magnitudes.  2MASS data for Hyades members from
Gizis et al. (1999) is also shown to illustrate the steepness of
of main sequence for for M0-M7 dwarfs, which may lead to
errors in the distance estimates for M7 dwarfs.  
\label{fig-jkmk}}
\end{figure}

Using the available parallaxes for late-type dwarfs 
\citep{m92,trgm95,t96,k99}, we
find (Figure~\ref{fig-jkmk}) the linear fit 
$M_K = 7.593 + 2.25 \times$ J-K$_s$
This fit is only valid for M7 and later dwarfs
over the color range $1.0 \le$ J-K$_S \lesssim 1.6$, and 
should be modified as more L dwarf parallaxes are measured
at USNO.  The observed scatter is $\sigma = 0.36$ magnitudes.   
The distances and tangential velocities derived using this 
estimate are listed in Table~\ref{table-fluxes}.  We caution,
however, that the distances derived for many of the M7 and M7.5 dwarfs
may be underestimated in this paper.  As can be seen in 
Figure~\ref{fig-jkmk}, the main sequence bends
sharply at spectral type M7.  Our selection of only targets
with J-K$_s\ge1.0$ tends to select M6's and M7's with overestimated
colors, while a spectral classification error of only 0.5 subclasses
from M6.5 to M7.0 leads to a large error in $M_K$.

\section{Stars of Special Interest\label{special}}

A few of the targets deserve special comment.  
Our search has identified one very nearby star and a number
of very high proper motion dwarfs.  
The M8.0 dwarf 2MASSW J0027559+221932 has a photometric parallax 
that places it within ten parsecs; given the uncertainties,
it may lie within the eight parsec sample.  
We note also that a number of dwarfs in Table~\ref{table-fluxes}
have motions greater than 1 arcsecond per year, but do not
appear in the LHS Catalog even though they are visible on
the POSS plates.  

Seven L dwarfs are part of our Sample BC.  The L5 dwarf 
2MASSW J1507476-162738, also in \citet{4ldwarfs}, was selected for
this project but does not lie in the Sample BC area.  
Additional observations and discussion of 2MASSI J0746425+200032, 
2MASSW J0036159+182110 and 2MASSW J1439283+192915 are 
given in \citet{4ldwarfs}.  
2MASSW J1439283+192915 is in the original K99 paper while
2MASSW J0036159+182110 was observed at Keck for the K00 paper.
2MASSW J1300425+191235 is particularly surprising:  it 
has the J-K$_s$ color of an M8 dwarf but has an L1 spectrum.
It will be discussed further in an additional
paper, but we note here that the estimated distance
and tangential velocity is based on the J-K$_s$ color
and should be viewed with great caution.  Nevertheless, it
apparently has a high velocity, and is likely to be old.  
While we report photometric
distance estimates only, most of these dwarfs are
on the USNO parallax program and accurate distances should
be forthcoming.

Since our selection is based upon photometry only, we 
are sensitive to wide binary pairs where the
2MASS observations of the secondaries are unaffected by the primaries.  
Two M dwarf secondaries that do not meet the spatial restrictions
and were specially observed have been reported in \citet{gl376b}.
One of the Sample C ultracool M dwarfs also appears to be 
a secondary.  
The M8.0 dwarf 2MASSW J2331016-040618 is 447 arcseconds west and
65 arcseconds south of the F8 dwarf HD 221356.  Our photometric
parallax of 26.3 parsecs for the M dwarf is consistent with Hipparcos 
trigonometric parallax of 26.24 parcsecs \citep{hipparcos},
as are the observed proper motions.  This is apparently a wide
binary system with separation of 0.057 parsecs.  The F8 primary
may provide a useful age and composition constraint on the
M dwarf.

Two of our sources have been previously identified 
as candidate Hyades members.  
LP 475-855 has been discussed as an Hyades candidate by
\citet{e93} although it was rejected as too bright by
\citet{lhd94}.  Our photometry supports the latter conclusion,
although conceivably it could be a foreground (escaping?) Hyades member
if it is an unresolved equal-mass binary.  Our initial observation 
of this star found it in a flare state, with H$\alpha$ equivalent width
of 40 \AA.  A 25 December 1998 Keck spectrum found H$\alpha$ of
only 7 \AA, which is the value we report in Table~\ref{table-data}.  
LP 415-20 (Bryja 262) has been extensively discussed as a Hyades
member and has been classified as an M6.5 dwarf \citep{bryja}.
The difference in spectral types is within our uncertainties, 
and we note that VO is visible in Bryja's plot of the spectrum.  
The poor distance estimate for this object is consistent with our
belief that our M7 distances based on J-K$_s$ colors
are unreliable and should be viewed with caution.

\section{Colors\label{colors}}

\begin{figure}
\epsscale{0.8}
\plotone{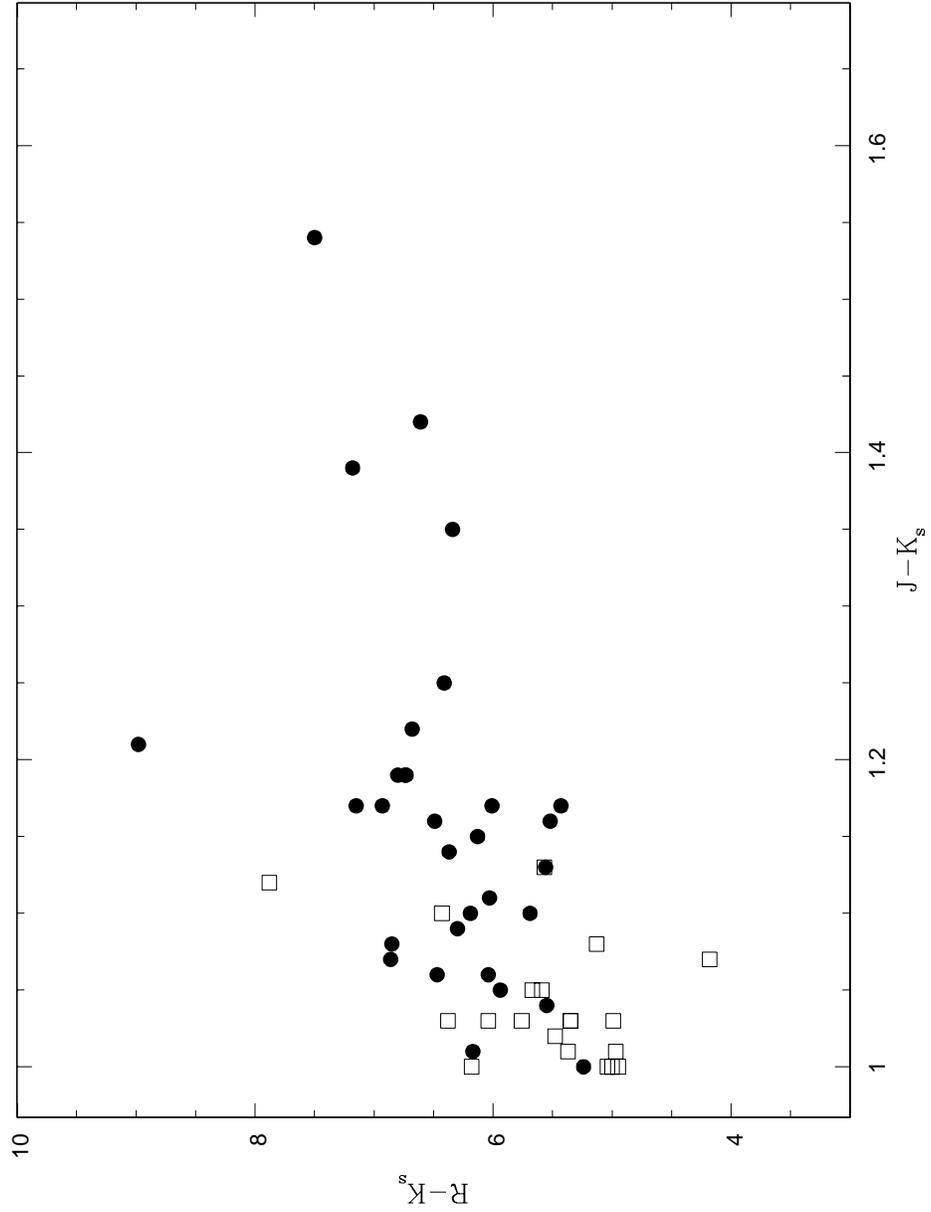}
\caption{Color-color diagram for the December sample using the
photographic R and 2MASS J and K$_s$ magnitudes.  M8 and later
dwarfs are solid circles, M7 dwarfs are open squares, and M6 and earlier
dwarfs are crosses.  A selection on R-K$_s >4.9$ would improve
our efficiency without excluding the ultracool M dwarfs.  
\label{fig-jkrk}}
\end{figure}

In Figure~\ref{fig-jkrk}, we plot the R-K$_s$,J-K$_s$ diagram for 
Sample B.  We concluded that adjusting our color criterion to  
R$_F$-K$_s > 4.9$ would increase our selection efficiency without
losing M8 and later dwarfs, and we adopted this selection criterion
for Sample C.  This observation is consistent with our estimate that
our simple $R_F$ zero-point is good to $\pm 0.5$ magnitudes.

\begin{figure}
\epsscale{0.8}
\plotone{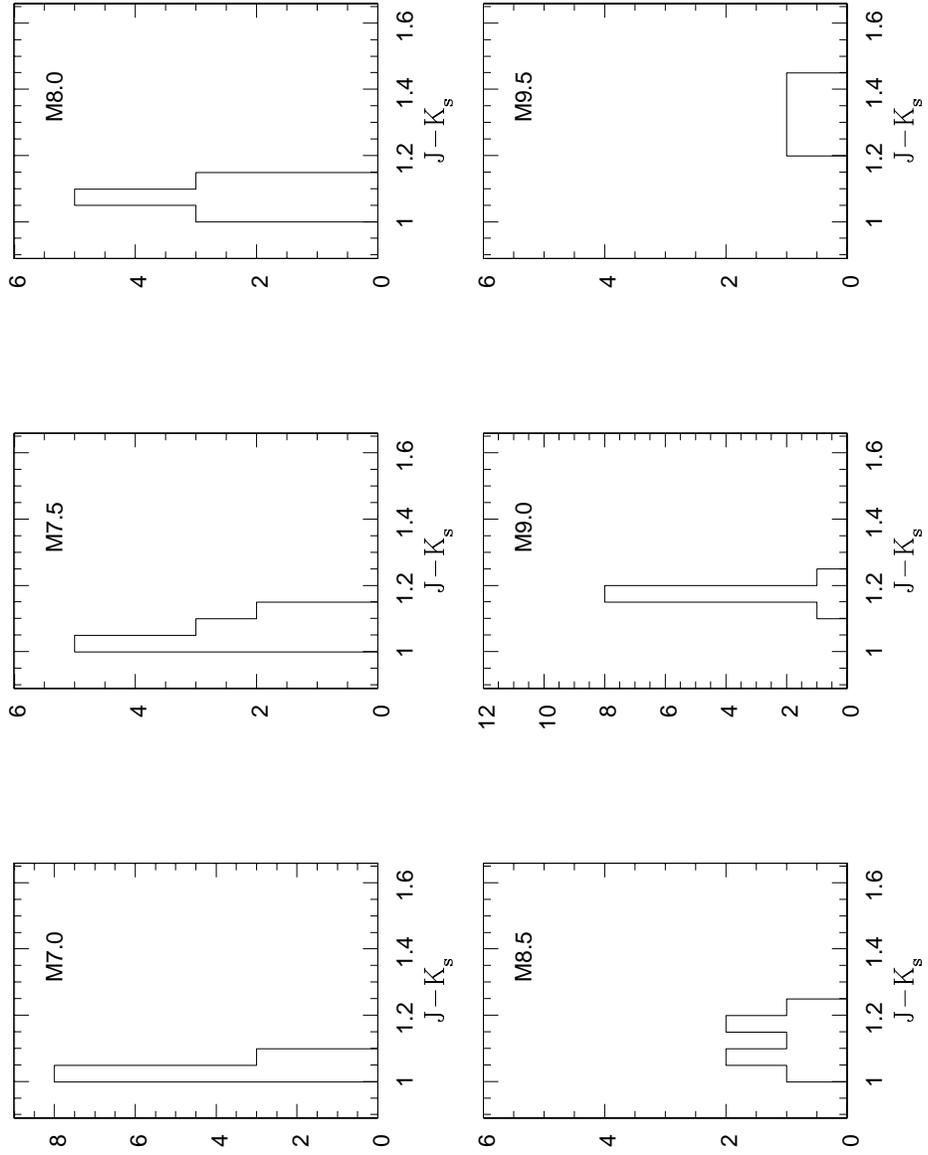}
\caption{2MASS J-K$_s$ color as a function of spectral type.  Each
bin is 0.05 magnitudes wide, approximately equal to the 2MASS uncertainty.   
Note the correlation between infrared color and far-red spectral type.
\label{fig-histjk}}
\end{figure}

Our observations show a good correlation between the far-red
spectral type and the 2MASS near-IR colors.  In Figure~\ref{fig-histjk},
we plot the observed J-K$_s$ color distribution as a function 
of spectral type.  As expected, the M7.0 and M7.5 distributions
are truncated by our requirement that J-K$_s \ge 1.0$.  
The histograms suggest that of order one M8-M8.5 dwarf may 
be expected to be missed due to this requirement, and it appears 
that it is very unlikely than any (normal) M9 dwarfs are missed.  
2MASS photometry for these bright sources is expected to be good
to 0.03 magnitudes.   

\begin{figure}
\epsscale{0.8}
\plotone{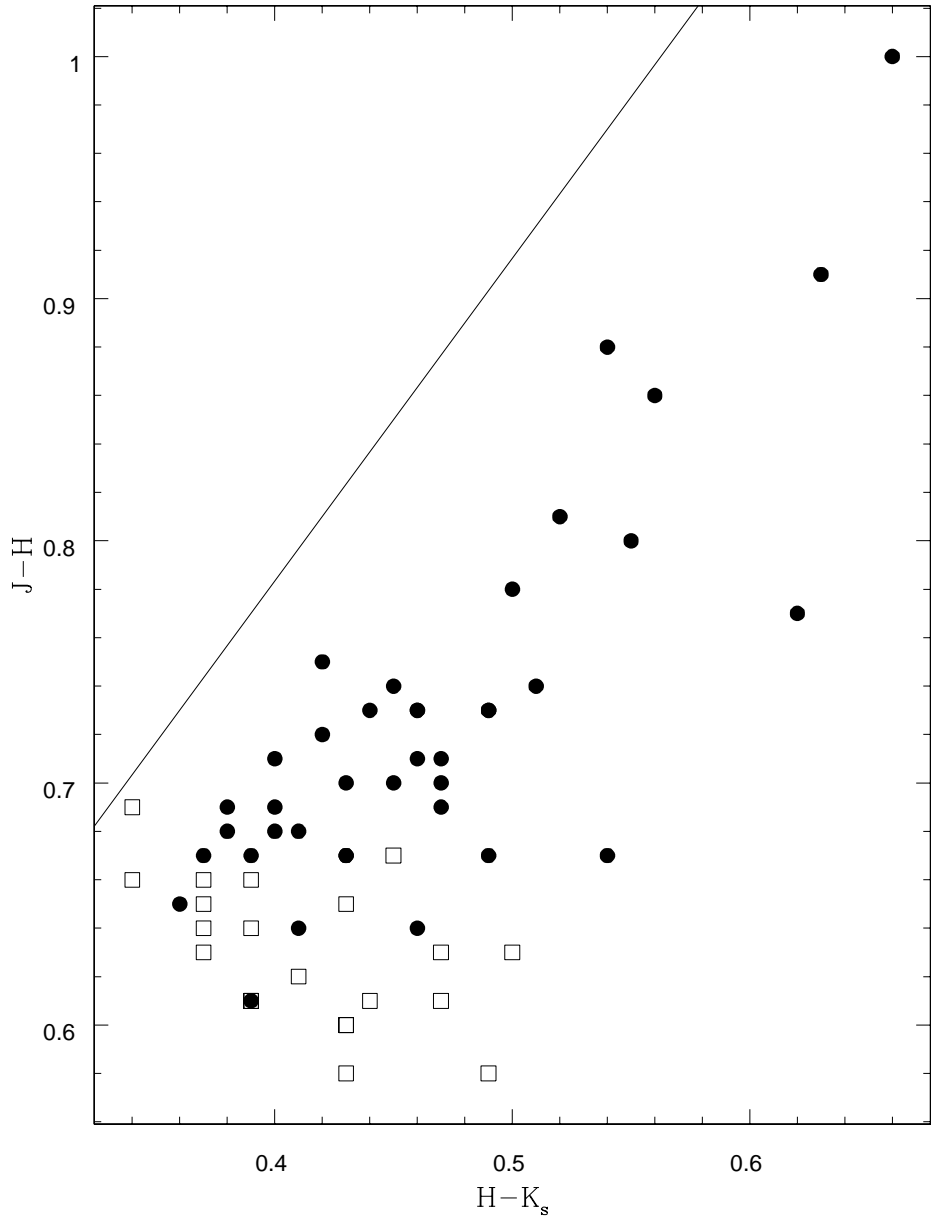}
\caption{2MASS near-infrared color-color diagram.  The solid lines
indicate our selection criterion. Note that all the M and L dwarfs
lie well below the line intended to exlude giants.   
\label{fig-jhhk}}
\end{figure}

In Figure~\ref{fig-jhhk}, we plot the 2MASS near-infrared color-color
diagram for our sample.  It is evident that the M/L dwarf
sequence lies well below our imposed J-H,H-K cut, and therefore
we are not missing ultracool M dwarfs due to this criteria.  
We fit the relation 
$J-K_s = (0.146 \pm 0.117) + (1.238 \pm 0.263) \times H-K_s$
assuming the errors in each color are 0.042 and including
the M7.0 to M9.5 dwarfs.  This relation may be convenient as
a representative sequence in the 2MASS color system.  
It is interesting to note that the dwarfs around 
(H-K,J-H) $= (0.45,0.62)$ are nearly all classified
as M7-M7.5 dwarfs, while the dwarfs at 
(H-K,J-H) $= (0.42,0.7)$ but with similar J-K$_s$
colors are nearly all classified as M8-M8.5 dwarfs.
This may reflect some relation between the
red-optical region (dominated by TiO and VO, and influenced by dust)
and the IR colors (dominated by H$_2$O and H$_2$, and also influenced by
dust) -- or it is due to some subtle bias in our classifications
or photometry (for example, perhaps we tend to select M7 dwarfs
whose H-K$_s$ color has been overestimated).   Note that one
of the outliers below the normal J-H,H-K$_s$ relation
is the peculiar L dwarf 2MASSW J1300425+191235.

\section{Luminosity Function\label{lf}}

Our Sample BC is the first large sample of bright, photometrically
selected ultracool M dwarfs.  
Using our data and derived distances, we can estimate the luminosity function
using Schmidt's (1968) $V/V_{max}$ technique.  The space density is 
$$\Phi = \sum \frac{1}{V_{max}}$$
$$V_{max} = \frac{\Omega}{3} \left( 10.0^{(K_{lim} - M_K + 5.0)/5.0}\right)^3$$
In our case, K$_{lim} = 12.0$ and $\Omega = 6040$ sq. degrees.
The corresponding variance is
$$\sigma_{\Phi}^2 = \sum \frac{1}{V_{max}^2}$$

The space densities are given in Table~\ref{table-lf}.  
We derive a space density of $0.0045 \pm 0.0008$ ultracool M dwarfs
per cubic parsec.  According to our adopted color-magnitude
relation, this is for dwarfs in the range $9.8 < M_K < 10.8$.
Therefore, the corresponding luminosity function bin
is $\Phi(M_K=10.3) = 0.0048 \pm 0.0009$ dwarfs per cubic
parsec per K magnitude.  Since \citet{tmr93} have
shown that $BC_K = 3.2$ for these late dwarfs, this may be
represented in bolometric magnitudes as
$\Phi(M_{bol}=13.5) = 0.0048 \pm 0.0009$
dwarfs per cubic parsec per bolometric magnitude.
We note, however, that this value excludes the M7 dwarfs which
will also contribute near M$_K = 9.8$, so our value is a lower limit.   
The space density
for the early L dwarfs is half that of the ultracool M dwarfs,
although we caution that the distance estimate for 2M1300
may be incorrect.  

Schmidt's statistic measures the uniformity of the density distribution
of a sample, effectively providing an estimate of sample completeness.
For a uniform sample,
$<V/V_{max}> = 0.5$ with an uncertainty of  $\frac{1}{\sqrt{12N}}$ where
N is the number of stars observed.  Both our L dwarf and
M8.0-M8.5 sample lie within 1 $\sigma$ of this value, suggesting
that we are complete.  The value for M9.0 to M9.5 dwarfs is 
more problematic, indicating that we have either excluded 
a few nearby, very bright M9 dwarfs, or that there happen to
be no such very nearby dwarfs in our survey volume.  

Our space density for the ultracool M dwarfs is 
consistent with Tinney's (1993) value of
$\Phi(M_{bol}=13.5) = 0.0076 \pm 0.0031$ dwarfs per cubic
parsec per bolometric magnitude, which was 
based on selection with R$_F$ and I$_N$ photographic magnitudes
but K followup of all VLM dwarfs to improve photometric parallaxes.
Only 6 dwarfs contributed to this bin, accounting for
Tinney's larger uncertainty relative to our sample.   
\citet{d99} have analyzed the DENIS Mini-survey
and found 19 M8 and later dwarfs, including 3 L dwarfs.
They do not estimate the M dwarf space density, but use the
three L dwarfs to estimate $\Phi(M_{bol}=15.3) \ge 0.011 \pm 0.006$
dwarfs per cubic parsec per bolometric magnitude.
We note that their estimated $M_K$ for the ultracool M dwarfs 
are inconsistent with
our adopted values (and Figure~\ref{fig-jkmk}) since they consider their
M8-M9 dwarfs to have $M_K>11$.  

Malmquist bias will affect our sample.  \citet{sip89}
have shown that the luminosity function will be overestimated by:
$$\frac{\Delta \Phi}{\Phi} = 0.5 \sigma^2 \left[ \left( 0.6 \ln 10 \right)^2
-1.2 \ln 10 \frac{\Phi^{\prime}}{\Phi} + \frac{\Phi^{\prime}}{\Phi^{\prime\prime}}\right]$$
A model luminosity function can be used to derive the first and second
derivatives.   
The scatter of parallax stars about the linear fit adopted here is
$\sigma=0.36$.  Adopting this value for $\sigma$, and making the
assumption that
the luminosity function is flat, we find that 
$\frac{\Delta \Phi}{\Phi} = 0.21$: i.e., the values we derive
overestimate the true space densities by $\sim 20\%$.  
Since this is only
a preliminary sample, we defer further analysis of the
Malquist bias, as well as the effect of unresolved binaries,
until additional data are available.  In the long term, trigonometric
parallaxes and searches for companions will allow this issue to
be addressed directly.

\begin{figure}
\epsscale{0.8}
\plotone{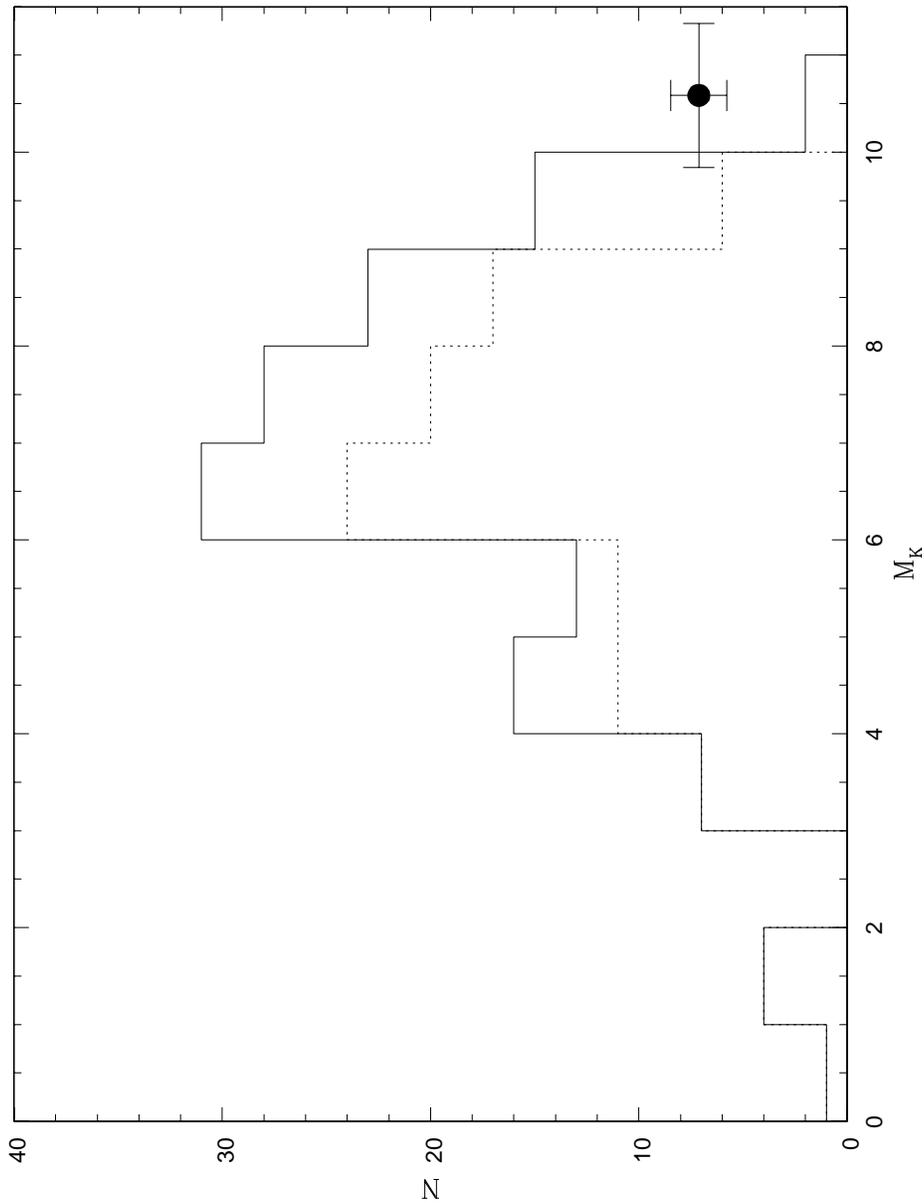}
\caption{Our observed space density of cool (M8.0-L4.5) dwarfs compared to the 
the Reid \& Gizis (1997) eight parsec sample (as updated in 
Reid et al. 1999).  The solid histogram count known secondaries,
while the dotted histogram excludes them.    
The steep dropoff at $M_K>10.0$ seen in both the
eight and 5.2 parsec samples is moderated by our space density.\label{fig-klf}}
\end{figure}

Our derived space density can best be compared to the luminosity function of
nearby stars.  Figure~\ref{fig-klf} plots the \citet{rg97a}
($\delta > -30\arcdeg$)
luminosity function for stars within eight parsecs 
with our M8.0-L4.5 data point added to it.  Our results suggest
that the dropoff seen in for the faintest ($M_K>10$) dwarfs in
the eight parsec sample is in part due to 
incompleteness.  
\footnote{Note that,  as seen in Reid \& Gizis's Fig. 2,
the oft-used 5.2 parsec
sample shows the same feature, albeit with less significance due
to the very small volume.}
Applying standard mass-luminosity relations to $\Phi(M_K)$ derived
from the 8-parsec sample implies a turnover in the mass function close to
the hydrogen-burning limit. There is, however, no reason to expect the
star formation process to be cognizant of the mass limit for hydrogen
burning.  \citet{reidmf} have modelled
a sample of twenty 2MASS and and three DENIS L dwarfs and 
conclude that the substellar mass function is consistent with an
extension of the power-law matched to data for stars with masses between
0.1 and $1 M_\odot$. The higher space densities measured for ultracool dwarfs
in this paper suggest a greater degree of continuity across the
stellar/substellar boundary.  
Continuation of the present survey
should identify the missing dwarfs within eight parsecs.  

If compared to the classical ``photometric'' luminosity functions
\citep{sip89,t93} which have a peak at $M_{bol}=10$ and a
dropoff to $M_{bol}=12$, then our data would imply a rise in the luminosity
function at the stellar/substellar boundary.  However, 
this peak and dropoff are an artifact of the data analysis due
to the incorrect assumption  of a linear color-magnitude relation
\citep{rg97a} and/or other systematic errors such as unresolved
binaries \citep{ktg93}.  We believe that the nearby-star sample
is a better comparison for our sample, and we emphasize again that
the M dwarfs here identified require follow-up trigonometric
parallax determinations and high-resolution imaging and radial
velocity searches for companions to produce
a definitive luminosity function.

\section{Activity and Kinematics\label{activity}}

\subsection{Review\label{review}}

The BC sample was not selected on the basis of proper motions, and
therefore is (relatively) unbiased in terms of kinematics.\footnote{
The existence of age-luminosity, age-metallicity,
metallicity-luminosity, age-activity, and age-kinematics correlations implies
that there may be kinematic and/or activity bias due to our 
Malquist-type luminosity bias.  If the distances are underestimated
due to bias, then the estimated tangential velocities will also
be biased.}  It is useful to review the 
properties of nearby disk stars and the already-known properties
of ultracool M dwarfs before discussing our kinematic and activity
measurements.  

Stars are born with low space velocity dispersions and 
and high chromospheric activity levels.  Over time, 
the space velocity of the stars increases as they interact
with the Galactic disk.  Using BAFGK dwarfs 
with known ages, \citet{w74} showed that the total
space velocity  increases from 
$\sigma_{tot}=19$ km/s at mean ages of 0.4 to $0.9 \times 10^9$ yr to
34 km/s at $2\times 10^9$ yr to $48$ km/s
at $5\times 10^9$ yrs. 
The high chromospheric activity levels of young
stars with convective envelopes is attributed to a dynamo 
which is driven by rotation.
As the star ages, angular momentum loss through the 
the stellar wind spins down the star, causing the
chromospheric activity to in turn decrease.    
\citet{w74} did
indeed find that \ion{Ca}{2} emission line strength in M dwarfs was related
to kinematics, with older stars showing less activity and
higher space velocities.  As 
progressively less massive (later spectral type) stars are considered,
the observed frequency of high H$\alpha$ activity increases.
This is {\it not} due only to the fact that H$\alpha$ emission is
more detectable against the cool photosphere -- 
\citet[hereafter HGR]{hgr96} showed
that the percentage of highly active M dwarfs increases with
cooler spectral types even when H$\alpha$ activity is compared
to the star's bolometric luminosity.  The increased lifetime of
activity is confirmed by observations of open clusters 
\citet{hrt99}.  The connection to rotation in early M dwarfs is confirmed 
by \citet{d98}, who have found that the incidence
of rapid rotators is higher among cooler M dwarfs, and that 
the rapid rotators are active.  There is some evidence that the
rotation-activity relation is breaking down \citep{hrg99}. 
In summary, mid-M dwarfs maintain H $\alpha$ emission for billions of years
as they slowly spin down.
Kinematics are a good age indicator, but only in a statistical sense --
individual stars can not be accurately dated by the velocities. 

For the ultracool M dwarfs, there is considerable evidence that
the standard stellar age-activity and rotation-activity relations 
no longer apply.  \citet{bm95} found that the M9.5
dwarf BRI 0021-0214 had very rapid rotation ($v \sin i = 40$ km/s) but 
little H$\alpha$ activity.  
TR have shown that the lithium M9 brown dwarf
LP 944-20 \citep{t98} is also a member of this class 
of ``inactive, rapid rotators''
as are the two of the DENIS L dwarfs \citep{mbdf97}.  
TR argue that observations of these field objects as well as 
open clusters indicate that the violation of the age-activity
connection is primarily correlated with mass (the physical
mechanisms remain unknown).
\citet{b99}  reports that rapid rotation is more common among
objects of lower luminosity, and proposes that the H$\alpha$ activity is
powered by a turbulent dynamo that is quenched at high rotation rates.  
There is some evidence that even among the ``inactive, rapid
rotators'' $\log \left({L_{H\alpha}}\over{L_{bol}}\right)$  
is related to age, as it decreases from $\sim -4.6$
for the Pleiades to $\sim -5.2$ for $\sim 0.5-1.0 \times 10^9$ yr
brown dwarfs.  This trend may be true for younger ages,
since the very low mass ($\sim 0.01-0.06 M_\odot$), very young ($<10$ Myr) 
M8.5 brown dwarf $\rho$ Oph 162349.8-242601 
has $EW_{H\alpha} > 50$\AA~ \citep{luhman,pc0025},
while the young ($\sim 1$ Myr),  possible brown dwarfs 
($M \approx 0.07 M_\odot$) V410 Tau X3 (M8.5) and X6 (M6) 
have $EW_{H\alpha} \approx15$ \AA~ \citep{pc0025}.  Both imply higher activity
levels than their older Pleiades and field counterparts,
though it should be noted that they are also lower mass.  
In contrast to the ``inactive, rapid rotators',''
some ultracool M dwarfs do show H$\alpha$ emission, but they
have lower rotation rates ($\lesssim 20$ km/s, TR).  
However, even a rotation rate of $\sim 5$ km/s is adequate to 
maintain H$\alpha$ emission in mid-M dwarfs \citep{b77},
so the limits on the ``low'' rotation rates of these ultracool M dwarfs
are not surprising by comparison. 
A so-far unique field object is the M9.5 dwarf PC 0025+0447
\citep{pc0025discover,pc0025}, whose
quiescent H$\alpha$ emission ($300$ \AA) is comparable to
highly active mid-M dwarfs in terms of
${L_{H\alpha}}\over{L_{bol}}$.  The nature of this object
and its emission is uncertain --- \citet{pc0025}
argue that this object is a very young brown dwarf,  suggesting that 
the USNO parallax indicating ordinary ultracool M dwarf luminosity 
is incorrect.  Some of the known M8-M9
dwarfs are definitely brown dwarfs.  The ``inactive, rapid rotator'' 
M9 dwarf LP 944-20 has lithium absorption and a luminosity that 
indicates it has a mass between 0.056 and 0.064 $M_\odot$ \citep{t98}.  
The Pleiades brown dwarfs Teide 1 and Calar 3 have spectral types of 
M8, lithium absorption, and anomalous VO and Na features due to
low-surface gravity \citep{mrzo96}.  

\subsection{The Properties of M and L dwarfs}

\begin{figure}
\epsscale{0.8}
\plotone{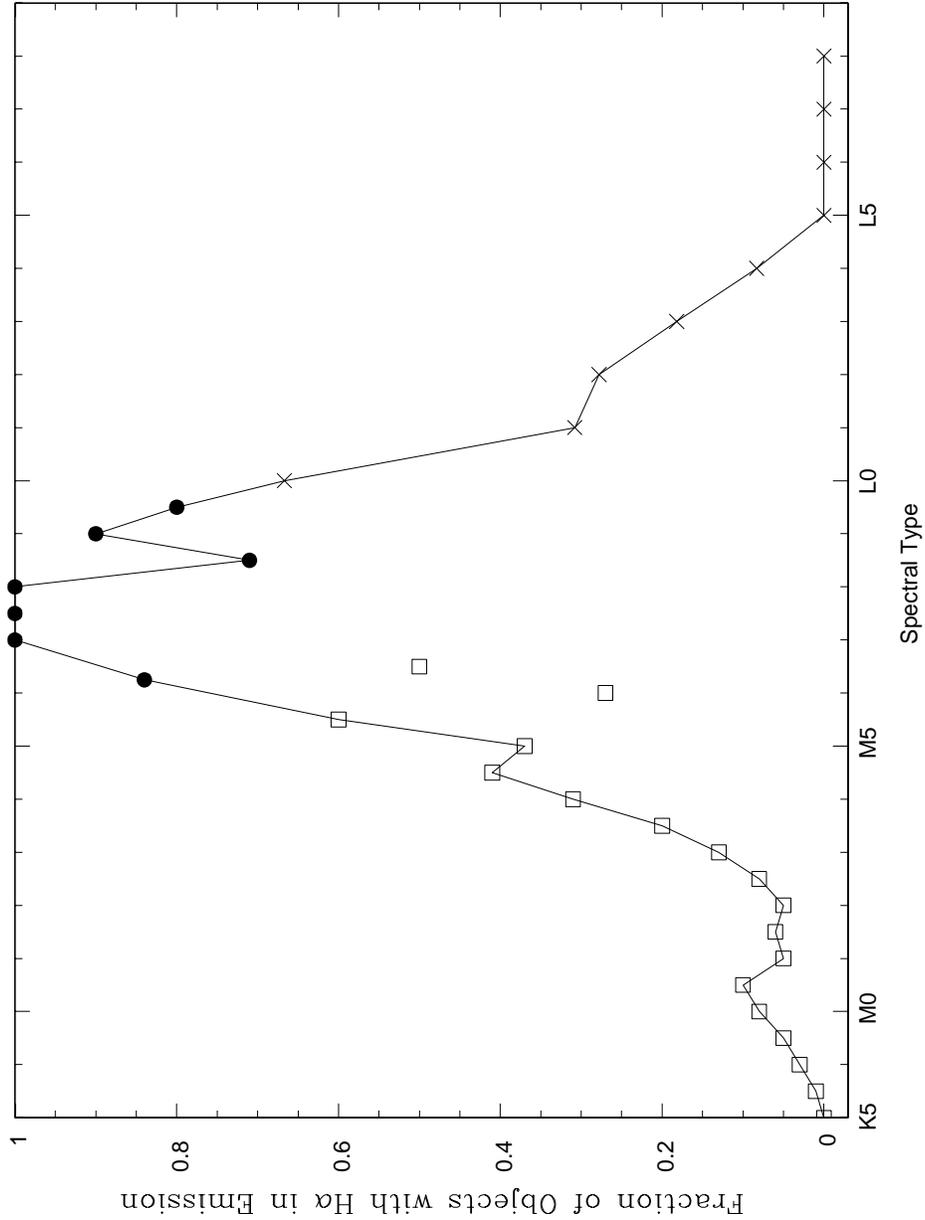}
\caption{The observed percentage of H$\alpha$ emission line dwarfs
amongst K5 to M6.5 dwarfs (HGR, open squares), M6 to M9.5 dwarfs (this paper,
solid circles), and L0 to L8 dwarfs (K99;K00, crosses).  
K5 and K7 dwarfs are plotted as -2 
and -1 respectively, while the L dwarfs are plotted with 10 added to
their subclass.  The HGR M6 and M6.5 dwarfs may be 
affected by kinematic bias, leading to an underestimate
of the number of emission stars.  The solid line connects the
three studies, adopting this paper's values for M6 dwarfs 
over the HGR values.    
\label{fig-percentha}}
\end{figure}

The BC sample of ultracool M dwarfs provides the first 
opportunity for a
thorough investigation of the distribution of activity in these 
VLM dwarfs.
In Figure~\ref{fig-percentha}, we compare the percentage of
ultracool M dwarfs observed in emission to the HGR statistics for
nearby K7 to M6.5 dwarfs.  Note that that $\sim 80$
to $\sim 300$ dwarfs contribute to each of the HGR bins
up to spectral type M4.5.  Only a few objects contribute to
the M6.0 and M6.5 bins, which also are probably   
kinematically biased against young, active stars due
to incompleteness in the pCNS3.\footnote{HGR's statistics for 
M7 and later dwarfs are sparse, but
further observations have revealed that they are incorrect.  Their Table
5 should show that 2 of 2 M7 dwarfs, 2 of 2 M8 dwarfs, and 
2 of 3 M9.0-M9.5 dwarfs show emission.} 
We extend our M dwarf sample
to even cooler dwarfs by using the K99 and K00 data, 
who report the strength of H$\alpha$ emission in their L dwarfs.  Their sample
is photometrically-selected and kinematically unbiased.
The sample shows a steady decline in H$\alpha$ emission frequency
from 60\% (80\% if a marginal detection is included) for type L0 down to
only 8\% (25\% if two marginal detections are included) for L4 dwarfs.
None of the twenty dwarfs with spectral type
L5.0 or later show definite emission (two have marginal detections).  
For the earlier dwarfs, emission
of 1 \AA~equivalent width would have been detected in almost all the
objects; however, the upper limit on the equivalent widths for
the latest L dwarfs was typically somwhat larger.  The 
fact that there is so little photospheric continuum for the latest
L dwarfs around the H$\alpha$ feature should compensate
for the lower sensitivity in terms in equivalent widths.

The data indicate that the frequency of emission increases 
with later spectral type (cooler temperatures), until at spectral 
type M7 all of our targets show detectable emission.  Indeed, 
we are not aware of {\it any} inactive M7 dwarfs 
(HGR; Gizis \& Reid 1997).  This indicates
that the dwarfs can maintain detectable levels of activity for the
lifetime of the Galactic disk.  Later than M7,
the H$\alpha$ emission frequency begins to decrease, with our sample
of ultracool M dwarfs merging cleanly with the L dwarfs.  This 
coincides with the breakdown of the rotation-activity 
relation already noted for M9 and L dwarfs and reflects 
the apparent relative inability of the ultracool M dwarfs to heat 
the chromosphere discussed by TR.   The percentage of emission at
spectral type M6 is particularly uncertain, as the HGR sample
may be biased toward higher velocity, hence older, stars 
at such low luminosities.  Sixteen of our nineteen M6-M6.5
dwarfs show emission,
but they have been effectively selected on the basis
of unusually red $J-K_s$ colors, and may be biased in some way.
In any case, there is little doubt that some high velocity,
presumably very old, M6 dwarfs are no longer active.

\begin{figure}
\epsscale{0.8}
\plotone{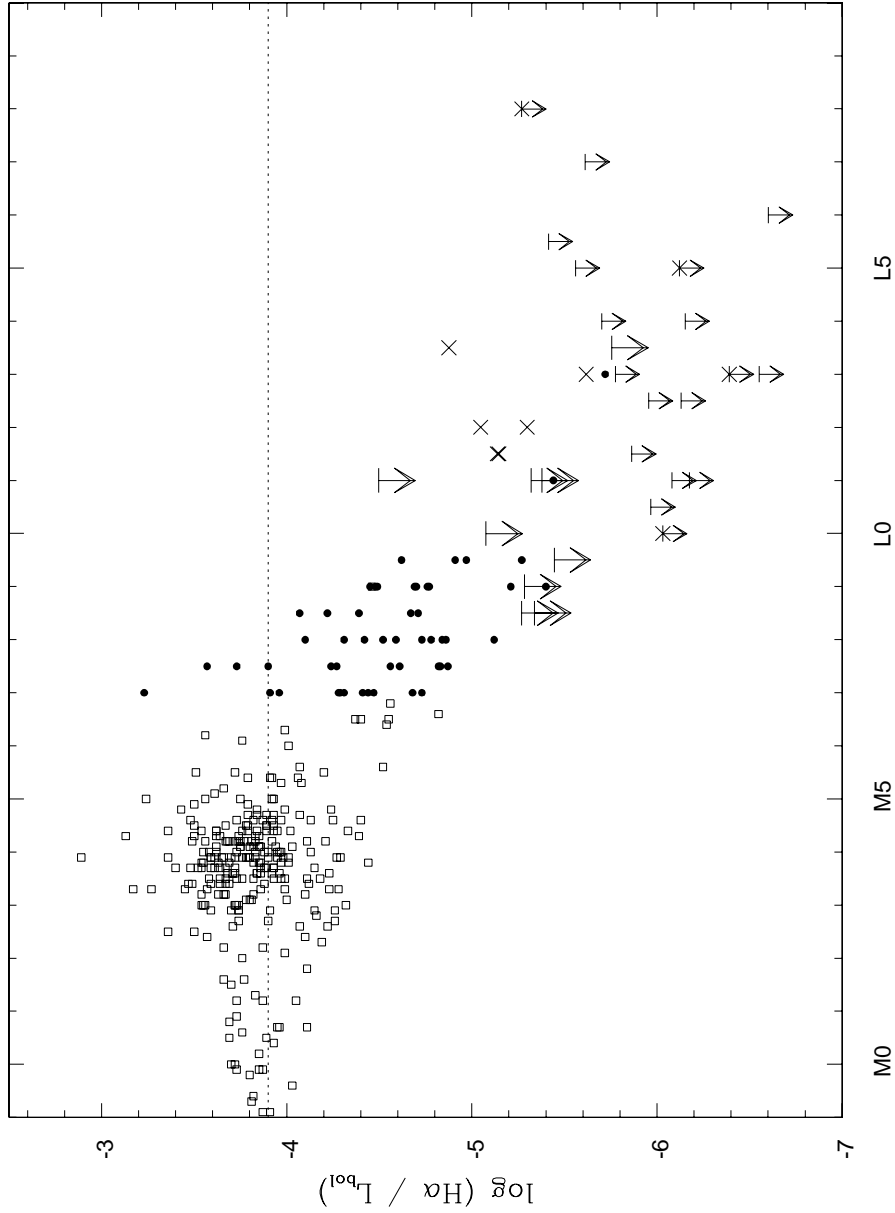}
\caption{The H$\alpha$ luminosity relative to the bolometric luminosity as a
function of spectral type for our ultracool M dwarfs (solid circles), the
earlier M dwarfs of HGR (open squares), and
the L dwarfs of K99 (crosses).  For our ultracool M dwarfs, approximate 
upper limits are plotted
assuming an H$\alpha$ equivalent width of $2$ \AA.  The dotted line
at -3.9 is the level at which any M dwarf would be observed in emission.
None of the M8 or later dwarfs have activity levels above the -3.9 level.     
\label{fig-sp-habol}}
\end{figure}

Since the H$\alpha$ line is seen against an increasingly
faint photosphere for these ultracool dwarfs, the 
$\log \left({L_{H\alpha}}\over{L_{bol}}\right)$ ratio is
more indicative of the true level of activity.  We plot 
our M dwarfs, the HGR early M dwarfs, and the K99 L dwarf data in 
Figure~\ref{fig-sp-habol}.  For the K99 data, we have measured
the continuum level off the observed spectra and assumed
$BC_K=3.33$ as measured by Tinney et al. (1993) for GD 165B to
convert the K99 equivalent widths to flux ratios.    
Note that the increase in maximum observed activity levels from K7 to
the  peak at M3-M5 reflects the increased lifetime of emission
for the lower mass stars.  Earlier M dwarfs with activity levels near 
-3 are known in young clusters, but do not have long enough lifetimes to appear
in the local sample.  The lower envelope of data points is set
by the fact that 
the minimum observable H$\alpha$ emission of 1\AA~equivalent width 
corresponds to a lower luminosity fraction in cooler dwarfs.
The addition of our data to the HGR data clearly indicates that
beyond M7 the level of activity is indeed declining.   This decline
continues for lower (L dwarf) temperatures.  The decline is 
quite steep -- in only three subclasses (M8 to L1) the activity
drops by one full dex.  \citet{mldwarf} note a ``slight trend toward 
decreasing emission in the L dwarfs'' -- our conclusions differ
due to our larger sample and their use of equivalent widths only.

While the quiescent H$\alpha$ chromospheric activity is declining,
our data suggest that flare activity is common in the
ultracool M dwarfs, and may be a significant contributor to
the activity energy budget.   
We summarize evidence for variability in Table~\ref{table-flares}.
Since these events represent a strong enhancement of the
H$\alpha$ line strength, we suggest that they may be flares.       
At least a few dwarfs apparently maintain strong quiescent emission -- 
our H$\alpha$ line strength of 29 \AA~ for LP412-31 is identical
with the value observed by \citet{mrzo96}.  Other ultracool
M dwarfs not in our sample have been seen to flare -- 
\citet{ourbri0021} recently 
observed a flare
on the ``inactive'' dwarf BRI 0021-0214, while 
\citet{mrm94} observed 
the H-alpha EW of LHS 2065 on two consecutive nights as 7.5 \AA~ 
then 20.3 \AA.  \citet{mldwarf} have observed flares in a number of
ultracool M dwarfs.  
Flaring activity has thus been observed in
ultracool M dwarfs with both very low and high levels of quiescent
emission.  

Assuming that these strong variations are due to flares, we 
estimate the flare rate from our own statistics.  At least four of the
fifty-three ultracool M targets were flaring the first time we observed
them for this program --- implying that ultracool M dwarfs spend
$\gtrsim 7\%$ of the time in a flare state. 
This is consistent with the observed flare rates of the
``inactive'' M9.5 dwarf BRI 0021-0214 \citep{ourbri0021} and
the monitoring of 2M0149 \citep{superflare}.   This flare rate
is a lower limit, since 
some of our other targets may also be flaring, but lacking 
additional spectra we cannot tell whether they are 
merely active as for LP 412-31, and since we have no way
of identifying weaker flares.   
The H$\alpha$ equivalent widths appear to be enhanced by
a factor of $\sim 10$ in the observed flares, implying that 
perhaps half of the H$\alpha$ luminosity is emitted
during flares.  

\begin{figure}
\epsscale{0.8}
\plotone{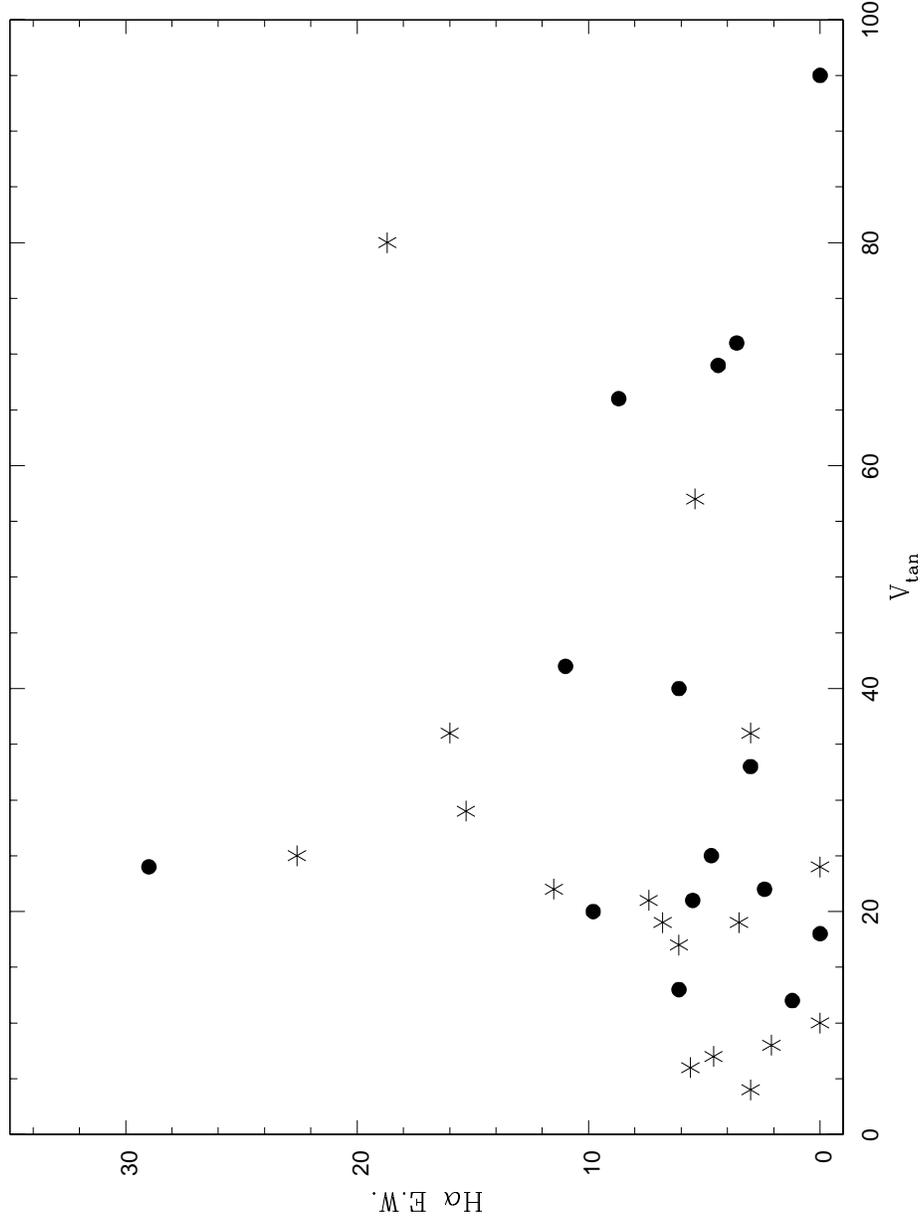}
\caption{H$\alpha$ equivalent width as a function of tangential velocity.
M8.0 - M8.5 dwarfs are plotted as stars and M9.0 - M9.5 dwarfs
are plotted as solid circles.  
All M8-M9 dwarfs with H$\alpha$ equivalent widths over 10\AA 
have large space velocities.  Note 2MASSW J0109216+294925, which has 
no emission but a high tangential velocity.   
\label{fig-vtanha}}
\end{figure}

We now consider the relationship between kinematics and
activity for the M8 - M9.5 dwarfs.  In Figure~\ref{fig-vtanha},
we plot the observed relation between the tangential velocity 
and H$\alpha$ emission.  There is a striking relationship
between activity and velocity, in a sense opposite to that 
observed in more massive M dwarfs.  All the dwarfs with
strong H$\alpha$ emission have large velocities ($v_{tan}>20$ km/s).
There is also a striking population of low-velocity, low-activity 
ultracool M dwarfs.  
With the exception of the high-velocity, inactive M9 dwarf 
2MASSW J0109216+294925, the least active stars appear to be drawn
from a lower velocity population.  It is difficult to 
fairly characterize the   
tangential velocity dispersion 
($\sigma_{tan}^2 = \sqrt{\sigma_{ra}^2 + \sigma_{dec}^2}$) 
of these populations, but the inactive, low-velocity
population in the lower left of Figure~\ref{fig-vtanha} 
may be characterized by $\sigma_{tan} \approx 15$ km/s.
While the low-velocity, low-activity population seems to 
to have EW$_{H\alpha} \lesssim 7$\AA~ and $v_{tan} \lesssim 25$ km/s,
we can calculate a velocity dispersion only for a purely
activity selected sample.  As an illustration, the
dwarfs with $EW_{H\alpha} <3$\AA~
(excluding 2MASSW J0109216+294925) have $\sigma_{tan} = 13$ km/s;
in contrast, those with more emission 
have  $\sigma_{tan} = 38$ km/s.
Using the approximation that 
$\sigma_{tot} = \sqrt{\frac{3}{2}} \sigma_{tan}$, the implied 
total space dispersions of
the two populations are 16 km/s and 47 km/s.  Comparing to \citet{w74},
the active M dwarfs are apparently drawn from a $\sim 5\times 10^9$
yr population, but the inactive M dwarfs are consistent with a
$\sim 0.5$ Gyr population.  
This estimate is crude at best,
but it seems clear that the overall ultracool M dwarf population is
drawn from a long-lived, presumably stellar population, while
the group of less active, low-velocity stars represent
a younger ($\lesssim 1$ Gyr) population .

Despite the smaller sample size, the properties of L dwarfs
are of considerable interest.
The Sample BC L0-L4 dwarf velocities
are typical of an old disk population.  
Even excluding 2M1300 due to unusual color, we find
$\sigma_{tan} = 56$ km/s, while including it we find
$\sigma_{tan}= 70$ km/s.  Only two of the Sample BC L dwarfs show
emission -- the low velocity 2M1108 has the strongest emission at
7.8\AA, while the high velocity 2M1506 has weak 1\AA~ emission.
While the velocities of only two L dwarfs are not definitive,
the velocity distribution of the inactive L dwarfs suggest they are
mostly old.  Adding the information provided by the work of K99 and K00 
to our data provides strong clues,
that just as in the ultracool M dwarfs, the traditional activity-age
relationship is broken, perhaps even reversed.  L dwarfs that
show lithium absorption are necessarily younger and lower mass
than L dwarfs of the same spectral type which have destroyed 
lithium.  Thus, using
the traditional stellar age-activity relation, one would expect them to be
more chromospherically active.  Consider the L1 to L4.5 dwarfs,
where lithium is detectable even at the low resolution of the
K99/K00 Keck LRIS observations.  Only one L dwarf, Kelu-1, shows both 
H$\alpha$ emission and lithium absorption.\footnote{It is interesting to 
note that \citet{b99} finds that Kelu-1 is rotating extremely rapidly:
80 km/s.}  Eleven other such L 
dwarfs show H$\alpha$ emission but do not have lithium absorption.
Twelve L dwarfs show lithium absorption but do not have 
H$\alpha$ emission (four of these have marginal H$\alpha$ detections or
noise consistent with emission of less than 2\AA).
While many L1-L4 dwarfs have neither H$\alpha$ emission nor
lithium absorption, it seems clear that the chromospherically
active L dwarfs are drawn from an older, more massive population
than the lithium L dwarfs.  
Beyond L4.5, there are no definite cases of H$\alpha$ emission, although
lithium absorption is present for $\sim 50\%$ of the L dwarfs.  

\begin{figure}
\epsscale{0.8}
\plotone{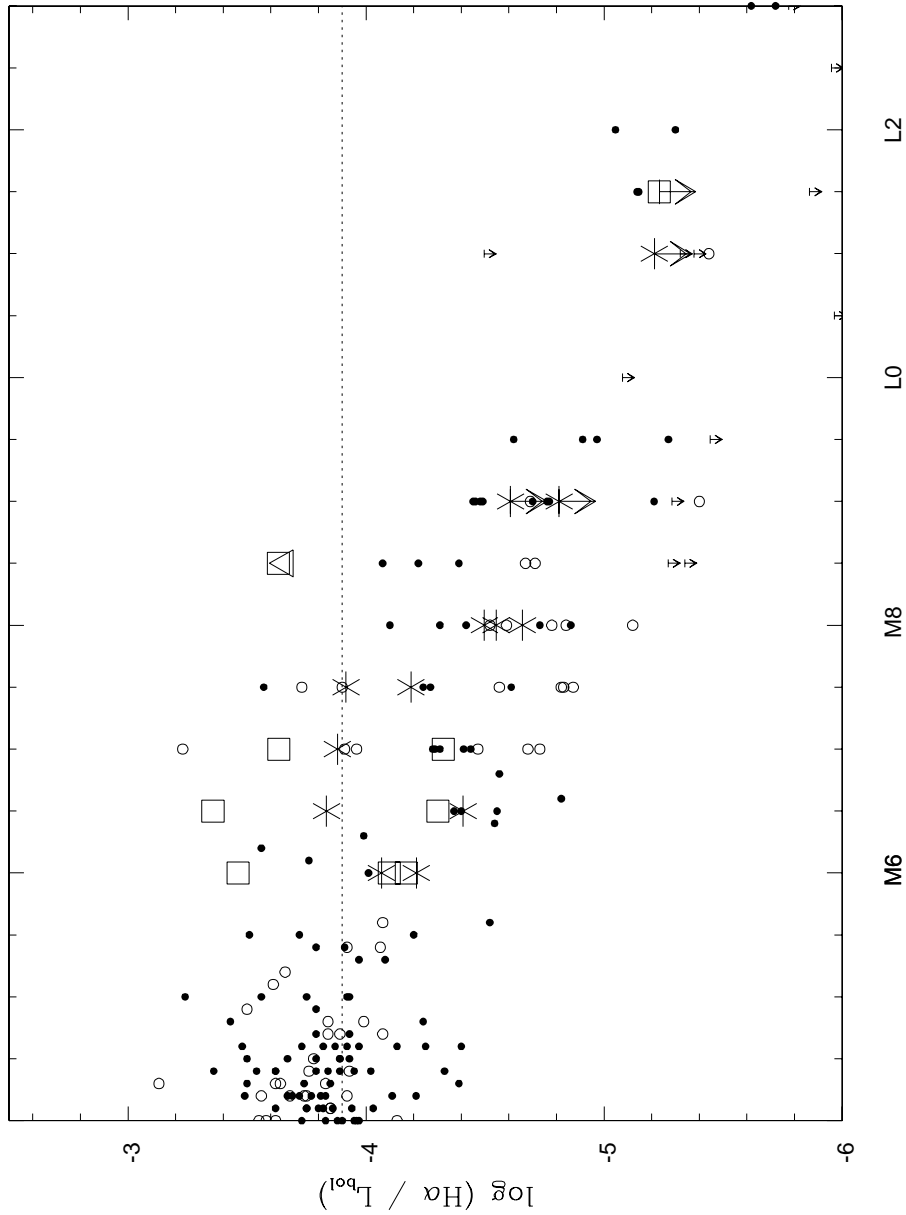}
\caption{The H$\alpha$ luminosity relative to the bolometric luminosity as a
function of spectral type for both cluster brown dwarfs and field
dwarfs.  Brown dwarfs from the $\sigma$ Ori cluster ($<10^7$ years),
$\rho$ Oph ($<10^7$ years), Pleiades ($\sim 10^8$ years) are shown
as open squares, open triangles, and six-pointed stars respectively.
Note that both cluster L dwarfs have only upper limits on the
detected H$\alpha$ emission.  The field M dwarfs are 
plotted as open circles if $v_{tan}< 20$ km/s and solid
circles for higher velocities.  
\label{fig-sphabol2}}
\end{figure}

Brown dwarfs have also been identified in nearby young clusters.
In Figure~\ref{fig-sphabol2}, we compare the activity of our field dwarfs to
young brown dwarfs.  Shown are young brown dwarfs from the 
$\sigma$ Ori cluster \citep{sigori1,sigori2} and the 
young brown in the $\rho$ Oph cloud \citep{luhman} -- both these
clusters are less than 10 Myr old.  Also shown are confirmed Plieades
brown dwarfs \citep{mrzo96,zopl,pl2,pl3} with age $\sim 10^8$ years.  
In order to suggest the age evolution of the field M dwarfs,
those with $v_{tan} < 20$ km/s have been marked as open symbols.  
These low-velocity dwarfs are likely to be younger than the other
field dwarfs.  It is evident that the Plieades
M8 and later brown dwarfs are not more active that the typical
field dwarfs, although the M6-M7 brown dwarfs appear to be more active.  
Like the young field L dwarfs that have lithium aborption,
the Pleiades and  $\sigma$ Ori L dwarfs do not show emission,
even though some older field L dwarfs do.  In the case of 
the $\rho$ Oph brown dwarf,
\citet{luhman} have shown that the emission is probably due to 
accretion from the circumstellar disk or envelope detected in
the mid-IR.  Similar accretion may account for the $\sigma$ Ori 
strong emitters, while the absence of accretion would account
for the weak emission in the other half of the $\sigma$ Ori brown 
dwarf sample.  If the emission is chromospheric, then 
some other factor (such as rotation) is needed to explain the 
large spread in activity levels.
It should be noted that these young brown dwarfs are 
probably hotter at a given spectral type \citep{luhman}, 
which suggests that if temperature were plotted the
brown dwarfs would appear even less active compared
to field dwarfs.  

We thus summarize the observations:
\begin{enumerate}
\item Although the fraction of dwarfs showing H$\alpha$ emission reaches 100\%
at spectral type M7, the fraction that show chromospheric 
activity drops rapidly for later spectral types.  
\item The fraction of energy in chromospheric H$\alpha$ for those dwarfs
that are active drops rapidly as a function of spectral type beyond M6.
\item Low velocity, kinematically young M8.0-M9.5 dwarfs have weaker
activity than many higher velocity, old M8.0-M9.5 dwarfs
\item Flaring is common among the M7-M9.5 dwarfs
\item The early L (L0-L4) dwarfs in Sample BC have old kinematics.
\item Early L (L1-L4.5) dwarfs with H$\alpha$ emission are old
and massive enough to have burned lithium 
\item L1-L4.5 dwarfs with lithium are unlikely to have H$\alpha$ emission.  
\item None of the L dwarfs later than L4.5 have H$\alpha$ emission 
but half have lithium.  
\item The two known L dwarfs in young clusters do not show H$\alpha$ emission
\item Young Pleiades M8-M9 dwarfs are less active than the higher velocity
field M dwarfs.

\end{enumerate}

\subsection{Discussion}

How can these observations be explained?   We believe they imply
that the maximum activity level is a strong function of 
temperature beyond spectral type M7, with lower temperature objects 
able to maintain less emission.  Additionally, beyond spectral
type M7 substellar dwarfs tend to have less activity than stellar dwarfs.

\begin{figure}
\plottwo{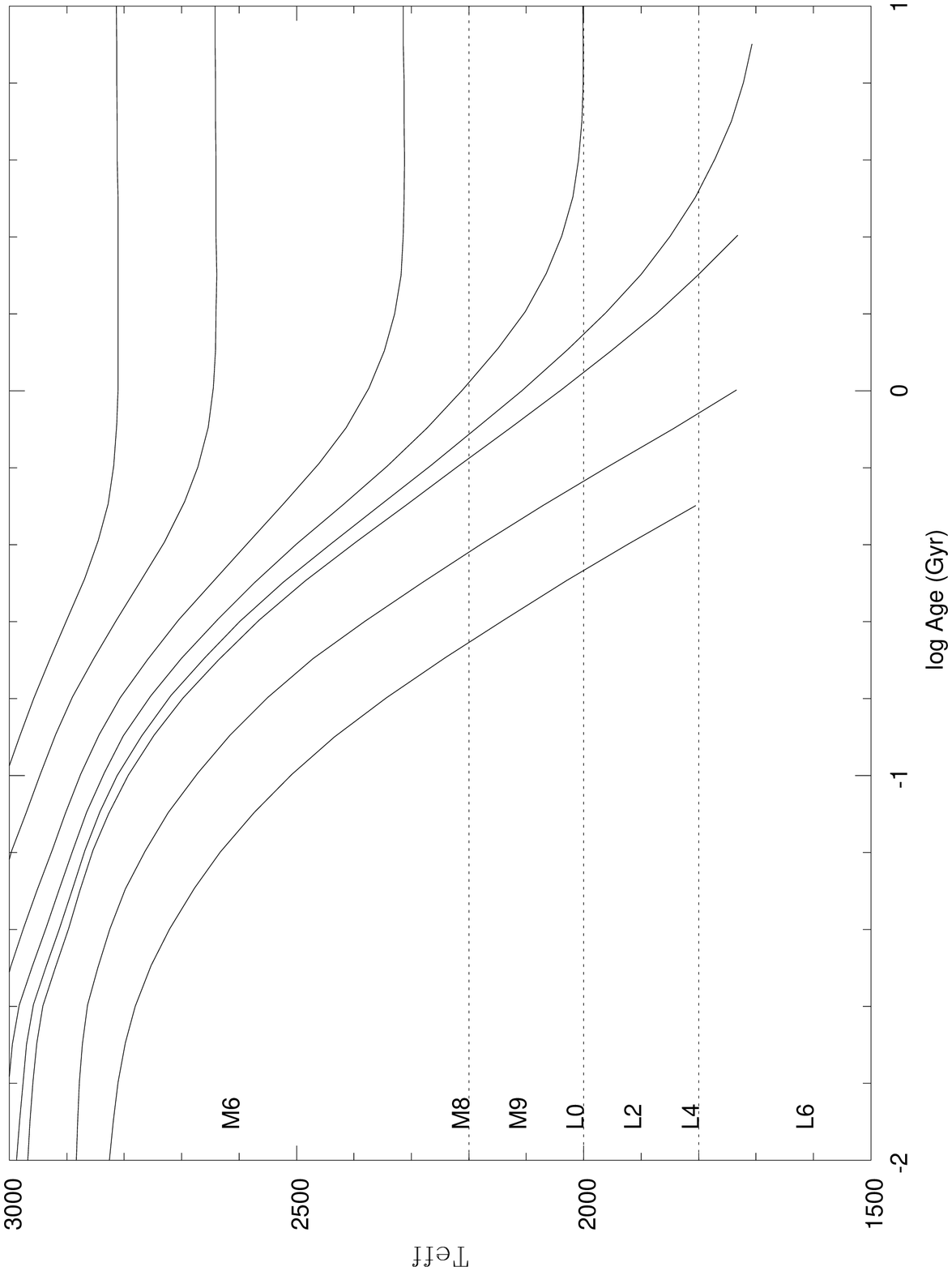}{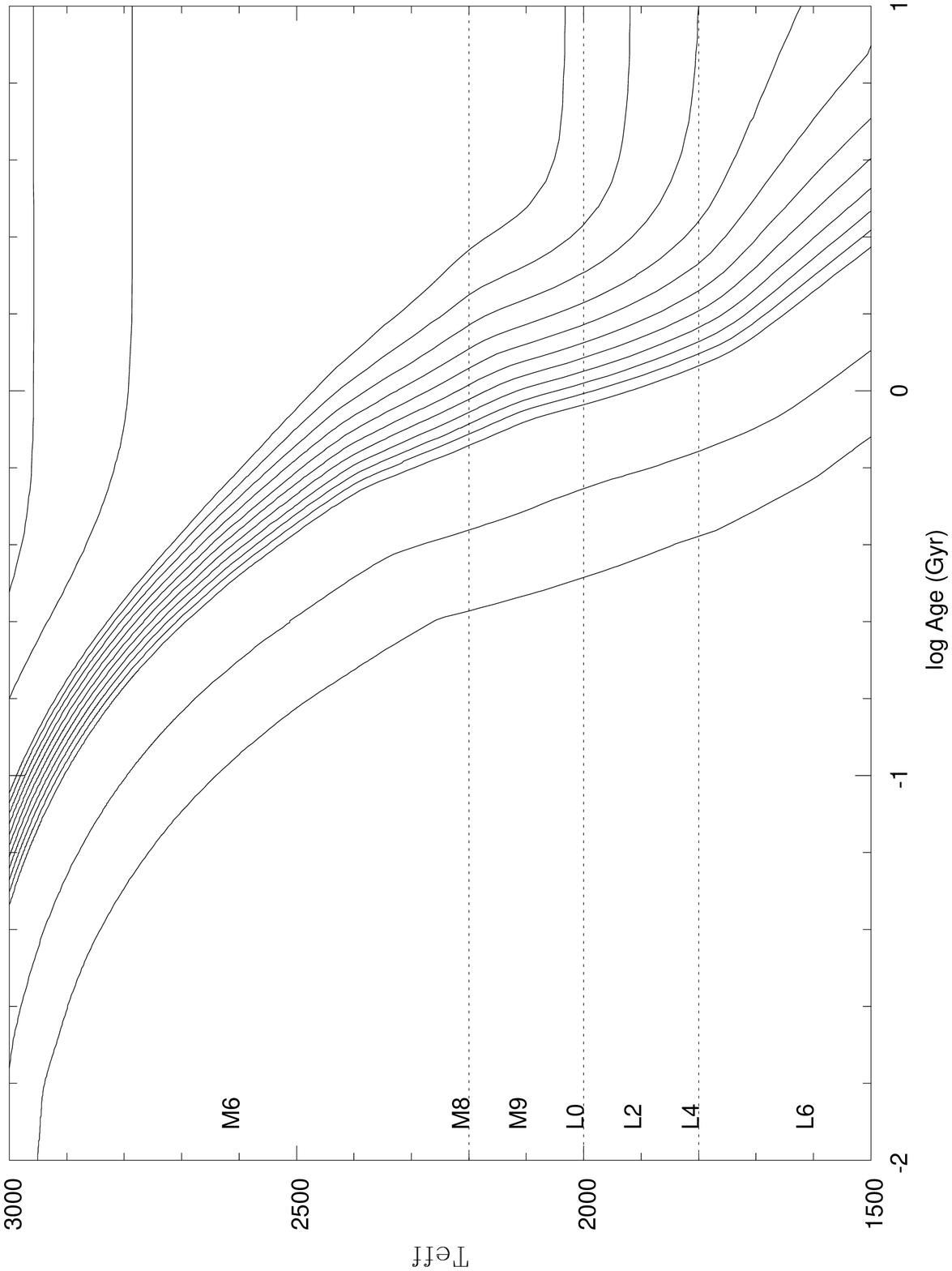}
\caption{Model calculations of brown dwarfs and the lowest mass stars by
Baraffe et al. (1998), including preliminary models for 
lower masses and younger ages (Baraffe \& Chabrier, priv. comm.)
and for Burrows et al. (1993, 1997).  
Along the left axis, the estimated
temperature scale of K99 and Reid et al. (1999) is indicated.  
The hydrogen burning limit is $0.072 M_\odot$ and the lithium
burning limit is $0.055 M_\odot$ \citep{cb97}.  
The models suggest that our spectral range will be populated
by stars with ${\rm age}\gtrsim 1$ Gyr, transition brown dwarfs burning
lithium with $0.4 \lesssim {\rm age} \lesssim 1$ Gyr, 
and brown dwarfs with lithium with age $\lesssim 0.4$ Gyr.
Both the model temperatures and the spectral-type temperatures are
uncertain.  These Baraffe et al. models use grainless model atmospheres.    
\label{fig-tage}}
\end{figure}

In Figure~\ref{fig-tage}, we plot evolutionary sequences from
\citet{burrows93} and \citet{bcah98}.  In the \citet{bcah98} models
the hydrogen burning limit is at 0.072 $M_\odot$ and the
lithium burning limit is at $0.055 M_\odot$.
Also shown is an estimated temperature scale from \citet{reidmf}
based on the arguments made in K99.  
The models indicate that it takes
$\gtrsim 10^9$ years for stars near the hydrogen burning limit to settle
into the M8 and cooler temperatures (Figure~\ref{fig-tage}).  
M8 and cooler temperatures are possible at younger ages, but
they are substellar objects which continue to cool with time.
Thus, the low velocity, low activity population is likely to 
be a population of substellar/transition objects.  By the time they are older
than 1 Gyr, they appear as L dwarfs or even cooler T dwarfs.  
Thus, the comparison of low velocity M8-M9.5 dwarfs to high
velocity M8-M9.5 dwarfs is the same as a comparison of younger, lower mass 
($\sim 0.07 M_\odot$) objects
to older, higher mass objects ($\sim 0.08 M_\odot$) at the same temperature.  
The same occurs when comparing the K99/K00 lithium L1-L4 
($\sim 0.055 M_\odot$) dwarfs
to the K99/K00 non-lithium L1-L4 ($\sim 0.07 M_\odot$) dwarfs.  
These mass estimates are only meant as illustrative values --
mass estimates are subject to many uncertainties and a range of
masses and ages will be sampled.  

Both the field M8-M9 dwarf and field L dwarf observations show that at a
given spectral type, 
the less massive dwarfs are less active, {\it even though they are younger.}
It is perhaps worth noting that theoretical models suggest that
the lower mass objects will be slightly more luminous with a
smaller surface gravity \citep{burrows93,bcah98}.  
The comparison to cluster brown dwarfs is consistent with this
trend (Figure~\ref{fig-sphabol2}).  
The Plieades brown dwarfs which have cooled to these
ultracool M and L temperatures are less active than old field
ultracool M and L dwarfs --- but more active than 
the low velocity ultracool M dwarfs.   There are thus strong suggestions,
as already noted in Section~\ref{review}, that activity levels
do decrease with age in brown dwarfs, and that therefore 
activity levels are dependent upon temperature, mass, 
and age.  The importance of accretion needs to be investigated
for the youngest ($<10$ Myr) ages.  The dominant effect
is temperature, as both young and old objects show the
rapid fall in activity levels beyond spectral type M7.

At the same time, the models suggest that stars, or at least very-long lived
hydrogen burning transition objects, are likely to exist down to L0 - L4
temperatures.  This is completely consistent with our empirical
finding that the early L dwarfs have old kinematics.  
We note that \citet{k99a} find a temperature of
$1900\pm 100$ K for the L4 dwarf GD 165B and constrain the age
to be greater than 1.2 Gyr using updated models and the white dwarf
primary's cooling age 
and argue it is just below the substellar limit, 
near the transition region between stars and brown dwarfs
(formally, they actually derive the minimum stellar mass using the models).  
L dwarfs with lithium must be below the lithium burning limit
($\sim 0.055 M_\odot$; Chabrier \& Baraffe 1997) and younger than 
1 Gyr (Figure~\ref{fig-tage}).   The inactivity of these
lithium L dwarfs demonstrates that the lowest mass objects cannot 
sustain significant activity at a temperature (or luminosity)
that is adequate for sustaining some activity in older but more
massive dwarfs.  

The decline in the frequency of activity may be associated
with two effects.  First, as later spectral types are considered,
a larger fraction of very low mass (substellar) objects 
contribute, and these are more likely to be inactive 
in field samples.  Second, activity among the L dwarfs may die
out in time, since the observed high velocity L dwarfs are
inactive -- although we cannot tell if they were ever active.
It would be of great interest to find whether or not the
early K99/K00 L dwarfs which are active have low or high 
velocities.  
The high-velocity, low-activity M9.5 dwarf 2MASSW J0109216+294925 may 
be a young brown dwarf that happens to have high velocity, 
an old object which has never been active, or 
an old stellar M9 dwarf whose chromospheric activity has declined
with age.  In any case, it is worthy of additional study.  
We note that the high observed flare rate implies that the 
rotation rate may decrease with time, even among the 
``inactive, rapid rotators'' if the flaring is associated with mass 
loss and/or a stellar wind.
We speculate this may provide a mechanism for the evolution of activity.
Additional observations are needed to determine what the
rotational velocities are as a function of mass, spectral type,
and age.  

What fraction of the ultracool M dwarfs are likely to be substellar?
While we cannot identify which individual objects are brown dwarfs, 
we can identify a number of probably young objects.  
Three of our M8.0-M9.5 dwarfs show no H$\alpha$ emission and
very low velocity; another three have equally low velocities
and $EW_{H\alpha}<3$\AA.  Out of a total population of 32 M8.0-M9.5
dwarfs, our data suggest that $10-20\%$ are brown dwarfs.  These objects
should be more likely to have lithium absorption (like LP 944-20),
but most of the brown dwarfs will be massive enough to burn lithium.  
Indeed, we note that none show the distinctive signs of
low surface gravity that characterize the Pleiades M8 brown dwarfs 
Teide 1 and Calar 3 \citep{mrzo96}, so none of our targets are
very young ($\sim 10^8$ years). 
Approximately ten objects belong to the low velocity, low activity
group in the lower left of Figure~\ref{fig-vtanha} -- that is a third
of the sample, but some fraction of these will stabilize
as hydrogen burning L dwarfs, in order to account for the
observed popoulation of high velocity early L dwarfs. 
These fractions will be somewhat overestimated for the Galactic 
disk population, since the old, large scale height population
will be underrepresented locally.  Other effects may also
be important, such as the fact that we have estimated
distances for all dwarfs using one color-absolute magnitude
relation.  Adding kinematic ages, as in this study, provides
an additional constraint on the modelling necessary to 
determine the field substellar mass function \citep{reidmf}.  

The nature of the ultracool M dwarfs has been debated for some time
in the literature.  While convential wisdom  
suggested that most if not
all field ultracool M dwarfs are stellar, many suggestions that most 
ultracool M dwarfs are substellar have been made, most of which have
been discredited.  We remark that \citet{b91} noted that 
the ultracool M dwarfs are expected to be a mixture of 
young brown dwarfs and older stars -- and he also noted that 
there was a paucity of high proper motion ultracool M dwarfs
expected from the stellar population 
in the LHS catalogue \citep{l79}. Our study has identified a 
number of high proper M dwarfs which appear on the red POSS plates
but were overlooked for the LHS catalogue -- evidently,
the faintest of these targets on the blue plate precluded
their detection by Luyten and  
contributed to the effect noted by Bessell.  
Our results show that most ultracool M dwarfs are old, and
hence stellar, but perhaps 10-20\% 
are a younger population of brown dwarfs.

We end our discussion with a few caveats due to our
photometric selection.  Both the relative numbers and kinematics of
``active'' and ``inactive'' M dwarfs will be changed if
one group is preferentially brighter at $M_K$ for its $J-K$ color.
Indeed, the inactive brown dwarf LP 944-20 lies one magnitude
below the active M dwarf LHS 2397a in the $J-K$, $M_K$ 
HR diagram.  Preliminary USNO parallaxes show instrinsic dispersion
in the HR diagram for the late M and L dwarfs (Dahn, private
communication).   If inactive ultracool M dwarfs are subluminous
compared to our adopted relation, we will have {\it overestimated}
their velocities; correspondingly, if the more active dwarfs
are ``superluminous,''  we have underestimated their velocities.  
Fortunately, this would only strengthen our evidence that 
active ultracool M dwarfs are older.    
The intrinsic dispersion presumably depends upon such ill-understood 
factors as metallicity, age, surface gravity, and dust formation.   
Another possible bias on activity levels is that we favor the inclusion
of unresolved binaries.  Amongst the earlier M dwarfs, very short period 
systems have enhanced chromospheric activity due to 
tidal effects maintaining high rotation rates\citep{ysh87}  --- 
however, even if this 
mechanism works in the ultracool M dwarfs, which seems unlikely
if the rotation-activity relation has broken down, only $\sim 5\%$ 
of earlier type M dwarfs show emission due to this effect, so 
it should be negligible.

\section{Summary\label{summary}}

We show that a sample of bright, nearby ultracool M and L dwarfs
can be selected without proper motion bias using 2MASS and PMM scans.  
Our initial samples include high proper motion objects, visible on 
the POSS plates, 
that should be added to an updated version of the  LHS Catalogue, and 
one M8.0 dwarf with
a photometric parallax that places it within 10 parsecs.  We intend
to continue this study in order to complete the nearby star catalog
for the lowest mass stars.  

Using our initial sample, we estimate the space density of dwarfs
near the hydrogen-burning limit.  We show that the dropoff near
the hydrogen burning limit in the five and eight parsec nearby star
samples is likely to be due to incompleteness. 
This is more consistent with a smooth relation across the
hydrogen burning limit.  Trigonometric parallaxes and searches for
companions will help improve the space density estimate.  

Most importantly, we use our spectroscopic observations of 
our well-defined sample to explore the relationships between 
age, kinematics, and chromospheric
activity for the ultracool M and L dwarfs.  We show that the observations
can be understood if activity is primarily related to temperature
and secondarily mass and age, and
that lower mass (substellar) objects have weaker chromosperes.   
Thus, the classical relation that strong H$\alpha$ emission implies
youth is not valid for these dwarfs.  Instead, strong H$\alpha$ 
emitters in the field are likely to be old ($\gtrsim 1$  Gyr) stars, while 
weaker emitters are often young ($<1$ Gyr), lower-mass brown dwarfs.  
This does not exclude the idea that for a given dwarf, H$\alpha$ activity
declines with age -- but spectral type (temperature) is the observable
in the field.
The local population of ultracool M dwarfs apparently consists both of
the most massive (lithium burning) brown dwarfs
and the lowest mass (hydrogen burning) stars, 
with the substellar objects making up a significant fraction of the sample.
The early L (L0-L4) dwarfs are consistent with an old, at least partially
stellar population.  The evidence thus suggests, as do some models, 
that early L dwarfs can be stable hydrogen-burning stars.  
Expansion of the sample with follow-up observations should clarify
the relative contribution of stars and brown dwarfs to these 
temperature ranges.

\acknowledgments

We thank Suzanne Hawley and Mike Skrutskie for useful discussions.
We thank the staffs of Las Campanas, Palomar, Keck, and Kitt Peak
observatories for their assistance in the observations and
the people at University of Massachusetts and IPAC for
their efforts in making 2MASS a reality.  
JEG and JDK acknowledge the
support of the Jet Propulsion Laboratory, California
Institute of Technology, which is operated under contract
with NASA.  
This work was funded in part by NASA grant AST-9317456
and JPL contract 960847.
INR, JDK, and JL acknowledge funding through
a NASA/JPL grant to 2MASS Core Project science
This publication makes use of data products from 2MASS, 
which is a joint project of the
University of Massachusetts and IPAC, funded by NASA and NSF.
Some of the data presented herein were obtained at the W.M. Keck
Observatory, which is operated as a scientific partnership among 
Caltech, the University of California 
and NASA. 
JEG accessed the DSS as a 
Guest User, Canadian Astronomy Data Centre, which is operated 
by the Herzberg Institute of Astrophysics, National Research Council 
of Canada. 
This research has made use of the Simbad database, operated at
CDS, Strasbourg, France.

\newpage

\begin{deluxetable}{lrrrrrlrc}
\tablewidth{0pc}
\tablenum{1}
\tablecaption{Targets}
\label{table-data}
\tablehead{
\colhead{Name} &
\colhead{RA (J2000)} &
\colhead{Dec}&
\colhead{J} &
\colhead{H} &
\colhead{K$_s$} & 
\colhead{Sp.} &
\colhead{H$\alpha$ EW}& 
\colhead{Sample}
}
\startdata
  2MASSI J0010325+171549  & 00:10:32.50 & +17:15:49.2 & 13.88 & 13.18 & 12.81 & M8.0  & 4.2 & A \\ 
  2MASSW J0016533+275534  & 00:16:53.37 & +27:55:34.9 & 12.82 & 12.12 & 11.77 & M6    & 7.2 & B \\ 
  2MASSW J0027559+221932\tablenotemark{a}  & 00:27:55.91 & +22:19:32.9 & 10.61 & 9.97 & 9.56 & M8.0  & 6.1 & B \\ 
  2MASSW J0036159+182110  & 00:36:15.98 & +18:21:10.2 & 12.44 & 11.56 & 11.02 & L3.5  & 0.0 & B \\ 
  2MASSI J0104377+145724  & 01:04:37.70 & +14:57:24.0 & 13.70 & 13.02 & 12.61 & M8.0  & 2.9 & A \\ 
  2MASSI J0105190+140740  & 01:05:19.02 & +14:07:40.9 & 13.59 & 12.92 & 12.55 & M7.0  & 10.2 & A \\ 
  2MASSW J0109216+294925  & 01:09:21.69 & +29:49:25.7 & 12.92 & 12.19 & 11.70 & M9.5  & 0.0 & B \\ 
  2MASSW J0130058+172143  & 01:30:05.82 & +17:21:43.8 & 13.66 & 12.98 & 12.58 & M8.0  & 0.6 & A \\ 
  2MASSW J0130144+271722  & 01:30:14.46 & +27:17:22.2 & 12.90 & 12.32 & 11.87 & M6    & 5.2 & B \\ 
  2MASSW J0140026+270150  & 01:40:02.64 & +27:01:50.6 & 12.51 & 11.82 & 11.44 & M8.5  & 0.0 & B \\ 
  2MASSI J0149089+295613  & 01:49:08.96 & +29:56:13.2 & 13.41 & 12.55 & 11.99 & M9.5  & 11.0 & B \\ 
  2MASSI J0218591+145116  & 02:18:59.13 & +14:51:16.2 & 14.18 & 13.58 & 13.25 & M7.0  & 6.0 & A \\ 
  2MASSI J0220181+241804  & 02:20:18.16 & +24:18:04.9 & 13.01 & 12.32 & 11.91 & M6    & 5.0 & B \\ 
  2MASSI J0240295+283257  & 02:40:29.51 & +28:32:57.6 & 12.62 & 11.99 & 11.62 & M7.5  & 8.4 & B \\ 
  2MASSI J0253202+271333  & 02:53:20.28 & +27:13:33.2 & 12.49 & 11.82 & 11.45 & M8.0  & 16.0 & B \\ 
  LP 412-31  & 03:20:59.65 & +18:54:23.3 & 11.74 & 11.04 & 10.57 & M9.0  & 29.0 & B \\ 
  2MASSI J0330050+240528\tablenotemark{b}  & 03:30:05.07 & +24:05:28.3 & 12.36 & 11.75 & 11.36 & M7.0  & 30.7 & B \\ 
  2MASSI J0335020+234235  & 03:35:02.08 & +23:42:35.6 & 12.26 & 11.65 & 11.26 & M8.5  & 4.6 & B \\ 
  2MASSW J0350573+181806\tablenotemark{c}  & 03:50:57.36 & +18:18:06.5 & 12.95 & 12.22 & 11.76 & M9.0  & 0.0 & B \\ 
  2MASSW J0354013+231633  & 03:54:01.34 & +23:16:33.9 & 13.12 & 12.42 & 11.97 & M8.5  & 6.8 & B \\ 
  LP 415-20  & 04:21:49.56 & +19:29:08.6 & 12.68 & 12.04 & 11.65 & M7.5  & 4.4 & B \\ 
  LP 475-855  & 04:29:02.83 & +13:37:59.2 & 12.67 & 11.98 & 11.64 & M7.0  & 40.5 & B \\ 
  2MASSI J0746425+200032  & 07:46:42.56 & +20:00:32.2 & 11.74 & 11.00 & 10.49 & L0.5    & 0.0 & B \\ 
  2MASSI J0810586+142039  & 08:10:58.65 & +14:20:39.1 & 12.71 & 12.04 & 11.61 & M9.0  & 6.1 & B \\ 
  2MASSI J0818580+233352  & 08:18:58.05 & +23:33:52.2 & 12.14 & 11.50 & 11.13 & M7.0  & 9.5 & B \\ 
  2MASSI J0925348+170441\tablenotemark{d}  & 09:25:34.85 & +17:04:41.5 & 12.60 & 11.99 & 11.60 & M7.0  & 4.7 & B \\ 
  2MASSW J0952219-192431  & 09:52:21.91 & -19:24:31.8 & 11.88 & 11.28 & 10.85 & M7.0  & 9.3 & B \\ 
  LHS 2243  & 10:16:34.70 & +27:51:49.8 & 11.95 & 11.29 & 10.95 & M7.5  & 43.8 & B \\ 
  2MASSI J1024099+181553  & 10:24:09.98 & +18:15:53.4 & 12.24 & 11.58 & 11.21 & M7.0  & 4.4 & B \\ 
  2MASSW J1049414+253852  & 10:49:41.44 & +25:38:52.9 & 12.40 & 11.75 & 11.39 & M6    & 6.9 & B \\ 
  2MASSW J1108307+683017  & 11:08:30.79 & +68:30:17.1 & 13.14 & 12.23 & 11.60 & L1    & 7.8 & C \\ 
  LHS 2397a  & 11:21:49.25 & -13:13:08.5 & 11.93 & 11.26 & 10.72 & M8.5  & 15.3 & B \\ 
  2MASSW J1127534+741107  & 11:27:53.48 & +74:11:07.9 & 13.06 & 12.37 & 11.97 & M8.0  & 3.0 & C \\ 
  2MASSW J1200329+204851  & 12:00:32.92 & +20:48:51.3 & 12.85 & 12.25 & 11.82 & M7.0  & 3.9 & C \\ 
  BRI 1222-1221  & 12:24:52.21 & -12:38:35.3 & 12.56 & 11.83 & 11.37 & M9.0  & 4.7 & B \\ 
  2MASSW J1237270-211748  & 12:37:27.05 & -21:17:48.1 & 12.67 & 12.05 & 11.64 & M6    & 8.3 & B \\ 
  LHS 2632  & 12:46:51.72 & +31:48:11.1 & 12.26 & 11.59 & 11.23 & M6.5  & 0.0 & C \\ 
  2MASSW J1300425+191235  & 13:00:42.55 & +19:12:35.6 & 12.71 & 12.07 & 11.61 & L1    & 0.0 & C \\ 
  2MASSW J1311391+803222  & 13:11:39.16 & +80:32:22.2 & 12.81 & 12.14 & 11.71 & M8.0  & 3.0 & C \\ 
  2MASSW J1336504+475131  & 13:36:50.46 & +47:51:31.9 & 12.64 & 12.06 & 11.63 & M7.0  & 5.0 & C \\ 
  2MASSW J1344582+771551  & 13:44:58.24 & +77:15:51.3 & 12.88 & 12.27 & 11.83 & M7.0  & 2.7 & C \\ 
  2MASSW J1403223+300755  & 14:03:22.34 & +30:07:55.0 & 12.69 & 12.01 & 11.63 & M8.5  & 18.7 & C \\ 
  2MASSW J1421314+182740  & 14:21:31.44 & +18:27:40.5 & 13.21 & 12.43 & 11.93 & M9.5  & 3.6 & C \\ 
  2MASSI J1426316+155701  & 14:26:31.61 & +15:57:01.3 & 12.87 & 12.18 & 11.71 & M9.0  & 1.2 & C \\ 
  2MASSW J1439283+192915  & 14:39:28.37 & +19:29:15.0 & 12.76 & 12.05 & 11.58 & L1    & 0.0 & C \\ 
  2MASSW J1444171+300214\tablenotemark{e}  & 14:44:17.17 & +30:02:14.3 & 11.68 & 10.97 & 10.57 & M8.0  & 7.4 & C \\ 
  2MASSW J1457396+451716  & 14:57:39.66 & +45:17:16.8 & 13.14 & 12.41 & 11.92 & M9.0  & 5.5 & C \\ 
  2MASSW J1506544+132106  & 15:06:54.40 & +13:21:06.0 & 13.41 & 12.41 & 11.75 & L3    & 1.0 & C \\ 
  2MASSW J1543581+320642\tablenotemark{f}  & 15:43:58.14 & +32:06:42.0 & 12.73 & 12.12 & 11.73 & M6.5  & 5.2 & C \\ 
  2MASSW J1546054+374946  & 15:46:05.40 & +37:49:46.1 & 12.44 & 11.79 & 11.42 & M7.5  & 10.9 & C \\ 
  2MASSW J1550381+304103  & 15:50:38.19 & +30:41:03.7 & 12.99 & 12.41 & 11.92 & M7.5  & 13.7 & C \\ 
  2MASSW J1551066+645704  & 15:51:06.63 & +64:57:04.6 & 12.87 & 12.15 & 11.73 & M8.5  & 11.5 & C \\ 
  2MASSW J1553199+140033  & 15:53:19.93 & +14:00:33.8 & 13.02 & 12.27 & 11.85 & M9.0  & 8.7 & C \\ 
  2MASSW J1627279+810507  & 16:27:27.93 & +81:05:07.9 & 13.04 & 12.33 & 11.87 & M9.0  & 6.1 & C \\ 
  2MASSW J1635192+422305  & 16:35:19.20 & +42:23:05.4 & 12.89 & 12.21 & 11.80 & M8.0  & 2.1 & C \\ 
  2MASSW J1658037+702701  & 16:58:03.77 & +70:27:01.7 & 13.31 & 12.54 & 11.92 & L1    & 0.0 & C \\ 
  2MASSW J1707183+643933  & 17:07:18.31 & +64:39:33.4 & 12.56 & 11.83 & 11.39 & M9.0  & 9.8 & C \\ 
  2MASSW J1714523+301941  & 17:14:52.34 & +30:19:41.0 & 12.94 & 12.28 & 11.89 & M6.5  & 5.4 & C \\ 
  2MASSW J1733189+463359  & 17:33:18.92 & +46:33:59.6 & 13.21 & 12.41 & 11.86 & M9.5  & 2.4 & C \\ 
  2MASSW J1750129+442404  & 17:50:12.90 & +44:24:04.5 & 12.79 & 12.17 & 11.76 & M7.5  & 2.7 & C \\ 
  2MASSW J1757154+704201\tablenotemark{g}  & 17:57:15.40 & +70:42:01.1 & 11.45 & 10.84 & 10.37 & M7.5  & 3.0 & C \\ 
  2MASSW J2013510-313651  & 20:13:51.02 & -31:36:51.3 & 12.67 & 12.06 & 11.67 & M6    & 6.2 & C \\ 
  LHS 3566  & 20:39:23.81 & -29:26:33.4 & 11.35 & 10.77 & 10.35 & M6    & 0.0 & C \\ 
  2MASSW J2049197-194432  & 20:49:19.74 & -19:44:32.5 & 12.87 & 12.24 & 11.77 & M7.5  & 13.1 & C \\ 
  2MASSW J2052086-231809\tablenotemark{h}  & 20:52:08.61 & -23:18:09.6 & 12.26 & 11.62 & 11.26 & M6.5  & 5.8 & C \\ 
  2MASSW J2113029-100941  & 21:13:02.94 & -10:09:41.0 & 12.86 & 12.22 & 11.81 & M6    & 0.0 & C \\ 
  2MASSW J2135146-315345  & 21:35:14.65 & -31:53:45.9 & 12.81 & 12.12 & 11.76 & M6    & 7.6 & C \\ 
  2MASSW J2140293+162518  & 21:40:29.32 & +16:25:18.4 & 12.94 & 12.27 & 11.78 & M8.5  & 0.0 & C \\ 
  2MASSI J2147436+143131  & 21:47:43.66 & +14:31:31.8 & 13.84 & 13.13 & 12.65 & M8.0  & 3.3 & A \\ 
  2MASSW J2147446-264406  & 21:47:44.62 & -26:44:06.6 & 13.04 & 12.37 & 11.92 & M7.5  & 3.9 & C \\ 
  2MASSW J2202112-110946\tablenotemark{i}  & 22:02:11.26 & -11:09:46.0 & 12.36 & 11.71 & 11.36 & M6.5  & 10.2 & C \\ 
  2MASSW J2206228-204705  & 22:06:22.80 & -20:47:05.8 & 12.43 & 11.75 & 11.35 & M8.0  & 5.6 & B \\ 
  2MASSI J2221531+115823  & 22:21:53.15 & +11:58:23.0 & 13.30 & 12.68 & 12.30 & M7.5  & 1.5 & A \\ 
  2MASSW J2221544+272907  & 22:21:54.43 & +27:29:07.5 & 12.52 & 11.92 & 11.52 & M6    & 3.6 & B \\ 
  2MASSW J2233478+354747\tablenotemark{j}  & 22:33:47.85 & +35:47:47.8 & 11.94 & 11.30 & 10.88 & M6    & 6.6 & C \\ 
  2MASSI J2234138+235956  & 22:34:13.88 & +23:59:56.1 & 13.14 & 12.33 & 11.81 & M9.5  & 4.4 & B \\ 
  2MASSI J2235490+184029\tablenotemark{k}  & 22:35:49.07 & +18:40:29.8 & 12.46 & 11.83 & 11.33 & M7.0  & 8.5 & B \\ 
  2MASSI J2255584+282246\tablenotemark{l}  & 22:55:58.45 & +28:22:46.7 & 12.55 & 11.94 & 11.54 & M6    & 5.2 & B \\ 
  2MASSW J2306292-050227  & 23:06:29.29 & -05:02:27.9 & 11.37 & 10.72 & 10.29 & M7.5  & 4.9 & C \\ 
  2MASSW J2313472+211729\tablenotemark{m}  & 23:13:47.29 & +21:17:29.5 & 11.43 & 10.75 & 10.42 & M6    & 6.0 & B \\ 
  2MASSW J2331016-040618  & 23:31:01.63 & -04:06:18.6 & 12.94 & 12.29 & 11.93 & M8.0  & 5.4 & C \\ 
  2MASSI J2334394+193304  & 23:34:39.44 & +19:33:04.2 & 12.77 & 12.07 & 11.64 & M8.0  & 22.6 & B \\ 
  2MASSI J2336439+215338\tablenotemark{n}  & 23:36:43.92 & +21:53:38.7 & 12.76 & 12.10 & 11.71 & M7.0  & 7.7 & B \\ 
  2MASSW J2347368+270206  & 23:47:36.80 & +27:02:06.8 & 13.19 & 12.45 & 12.00 & M9.0  & 3.0 & B \\ 
  2MASSW J2349489+122438\tablenotemark{o}  & 23:49:48.99 & +12:24:38.8 & 12.62 & 11.95 & 11.56 & M8.0  & 3.5 & C \\ 
  2MASSW J2358290+270205\tablenotemark{p}  & 23:58:29.00 & +27:02:05.5 & 12.71 & 12.05 & 11.68 & M6    & 5.9 & B \\ 
\enddata
\tablenotetext{a}{LP 349-25}
\tablenotetext{b}{LP 356-770}
\tablenotetext{c}{LP 413-53}
\tablenotetext{d}{LP 427-38}
\tablenotetext{e}{LP 326-21}
\tablenotetext{f}{LP 328-36}
\tablenotetext{g}{LP 44-162}
\tablenotetext{h}{LP 872-22}
\tablenotetext{i}{LP 759-17}
\tablenotetext{j}{LP 288-31}
\tablenotetext{k}{LP 460-44}
\tablenotetext{l}{LP 345-18}
\tablenotetext{m}{LP 461-11}
\tablenotetext{n}{LP 402-58}
\tablenotetext{o}{LP 523-55}
\tablenotetext{p}{LP 348-11}
\end{deluxetable}

\begin{deluxetable}{lrcclcc}
\tablewidth{0pc}
\tablenum{2}
\tablecaption{L Dwarf Data}
\label{table-ldwarfs}
\tablehead{
\colhead{Name} &
\colhead{CrH-a} &
\colhead{Rb-b/TiO-b} &
\colhead{Cs-a/VO-b} & 
\colhead{K I fit} &
\colhead{Type} 
}
\startdata
2MASSW J1108307+683017 &  1.27(0-1) &  0.88(1) &   0.84(1)   &  (0)  &   L1 V \\
2MASSW J1300425+191235 &  1.53(2)   &  0.81(1) &   0.81(0-1) &  (1)  &   L1 V \\
2MASSW J1506544+132106 &  1.44(1-2) &  1.18(3) &   1.13(3)   &  (3)  &   L3 V \\
2MASSW J1658037+702701 &  1.26(0-1) &  0.79(1) &   0.81(0-1) &  (2)  &   L1 V \\
\enddata
\end{deluxetable}

\begin{deluxetable}{lrrrrr}
\tablewidth{0pc}
\tablenum{3}
\tablecaption{Kinematics and Activity}
\label{table-fluxes}
\tablehead{
\colhead{Name} &
\colhead{d$_{phot}$} &
\colhead{$\mu_{\alpha}$} &
\colhead{$\mu_{\delta}$} & 
\colhead{V$_{tan}$} &
\colhead{$\log \frac{L_{H\alpha}}{L_{bol}}$} 
}
\startdata
  BR 1222-1221  & 16.6 & -0.262 & -0.187 & 25 & -4.70 \\ 
  LHS 2243  & 16.6 & -0.158 & -0.461 & 38 & -3.57 \\ 
  LHS 2397a  & 12.0 & -0.509 & -0.081 & 29 & -4.22 \\ 
  LP 412-31  & 11.7 & 0.349 & -0.251 & 24 & -4.45 \\ 
  LP 415-20  & 22.3 & 0.127 & -0.036 & 14 & -4.56 \\ 
  LP 475-855  & 22.2 & 0.103 & -0.016 & 11 & -3.23 \\ 
  2MASSW J0027559+221932  & 8.3 & 0.403 & -0.172 & 17 & -4.52 \\ 
  2MASSW J0036159+182110  & 11.1 & 0.837 & 0.104 & 44 & \nodata \\ 
  2MASSW J0109216+294925  & 18.7 & 1.014 & 0.348 & 95 & \nodata \\ 
  2MASSW J0140026+270150  & 19.4 & 0.061 & -0.252 & 24 & \nodata \\ 
  2MASSI J0149089+295613  & 17.4 & 0.207 & -0.466 & 42 & -4.62 \\ 
  2MASSI J0240295+283257  & 22.7 & 0.046 & -0.192 & 21 & -4.27 \\ 
  2MASSI J0253202+271333  & 20.1 & 0.370 & 0.088 & 36 & -4.31 \\ 
  2MASSI J0330050+240528  & 20.1 & 0.185 & -0.039 & 18 & -3.91 \\ 
  2MASSI J0335020+234235  & 19.2 & 0.058 & -0.043 & 7 & -4.71 \\ 
  2MASSW J0350573+181806  & 19.9 & 0.189 & -0.049 & 18 & \nodata \\ 
  2MASSW J0354013+231633  & 22.8 & -0.168 & 0.064 & 19 & -4.67 \\ 
  2MASSI J0746425+200032  & 10.4 & -0.358 & -0.054 & 18 & \nodata \\ 
  2MASSI J0810586+142039  & 20.3 & -0.034 & -0.128 & 13 & -4.69 \\ 
  2MASSI J0818580+233352  & 17.9 & -0.275 & -0.305 & 35 & -4.31 \\ 
  2MASSI J0925348+170441  & 22.5 & -0.232 & 0.010 & 25 & -4.28 \\ 
  2MASSW J0952219-192431  & 15.4 & -0.077 & -0.104 & 9 & -3.96 \\ 
  2MASSI J1024099+181553  & 18.2 & -0.144 & -0.070 & 14 & -4.73 \\ 
  2MASSW J1108307+683017  & 12.8 & -0.226 & -0.194 & 18 & -5.44 \\ 
  2MASSW J1127534+741107  & 24.3 & -0.016 & -0.030 & 4 & -4.84 \\ 
  2MASSW J1200329+204851  & 24.1 & -0.159 & 0.232 & 32 & -4.44 \\ 
  2MASSW J1300425+191235  & 20.3 & -0.789 & -1.238 & 141 & \nodata \\ 
  2MASSW J1311391+803222  & 21.3 & -0.068 & -0.348 & 36 & -4.86 \\ 
  2MASSW J1336504+475131  & 22.5 & 0.111 & -0.016 & 12 & -4.47 \\ 
  2MASSW J1344582+771551  & 23.7 & 0.072 & -0.005 & 8 & -4.68 \\ 
  2MASSW J1403223+300755  & 21.4 & -0.788 & 0.042 & 80 & -4.07 \\ 
  2MASSW J1421314+182740  & 19.6 & -0.744 & -0.182 & 71 & -4.97 \\ 
  2MASSI J1426316+155701  & 20.0 & 0.108 & -0.056 & 12 & -5.40 \\ 
  2MASSW J1439283+192915  & 18.5 & -1.245 & 0.392 & 114 & \nodata \\ 
  2MASSW J1444171+300214  & 12.5 & -0.101 & -0.336 & 21 & -4.42 \\ 
  2MASSW J1457396+451716  & 20.7 & -0.191 & 0.100 & 21 & -4.76 \\ 
  2MASSW J1506544+132106  & 12.1 & -1.092 & 0.001 & 63 & -5.72 \\ 
  2MASSW J1546054+374946  & 20.2 & -0.020 & -0.120 & 12 & -3.90 \\ 
  2MASSW J1550381+304103  & 24.2 & -0.112 & 0.107 & 18 & -3.73 \\ 
  2MASSW J1551066+645704  & 20.6 & -0.220 & 0.010 & 22 & -4.39 \\ 
  2MASSW J1553199+140033  & 21.1 & -0.659 & 0.072 & 66 & -4.48 \\ 
  2MASSW J1627279+810507  & 21.3 & -0.209 & 0.338 & 40 & -4.77 \\ 
  2MASSW J1635192+422305  & 22.4 & -0.073 & -0.010 & 8 & -5.12 \\ 
  2MASSW J1658037+702701  & 17.4 & -0.136 & -0.315 & 28 & \nodata \\ 
  2MASSW J1707183+643933  & 17.1 & 0.226 & -0.091 & 20 & -4.49 \\ 
  2MASSW J1733189+463359  & 17.6 & 0.044 & -0.257 & 22 & -5.27 \\ 
  2MASSW J1750129+442404  & 23.4 & -0.018 & 0.151 & 17 & -4.87 \\ 
  2MASSW J1757154+704201  & 11.7 & 0.006 & 0.338 & 19 & -4.83 \\ 
  2MASSW J2049197-194432  & 21.9 & 0.193 & -0.260 & 34 & -4.24 \\ 
  2MASSW J2140293+162518  & 20.7 & -0.008 & -0.102 & 10 & \nodata \\ 
  2MASSW J2147446-264406  & 22.9 & -0.054 & -0.232 & 26 & -5.04 \\ 
  2MASSW J2206228-204705  & 18.4 & 0.001 & -0.065 & 6 & -4.59 \\ 
  2MASSI J2234138+235956  & 17.6 & 0.829 & -0.034 & 69 & -4.91 \\ 
  2MASSI J2235490+184029  & 17.3 & 0.326 & 0.042 & 27 & -4.41 \\ 
  2MASSW J2306292-050227  & 11.3 & 0.889 & -0.420 & 53 & -4.61 \\ 
  2MASSW J2331016-040618  & 26.3 & 0.401 & -0.231 & 58 & -4.72 \\ 
  2MASSI J2334394+193304  & 20.0 & -0.236 & -0.117 & 25 & -4.10 \\ 
  2MASSI J2336439+215338  & 22.4 & 0.379 & 0.024 & 40 & -4.29 \\ 
  2MASSW J2347368+270206  & 22.2 & 0.313 & 0.033 & 33 & -5.21 \\ 
  2MASSW J2349489+122438  & 20.7 & 0.025 & -0.189 & 19 & -4.78 \\ 
\enddata
\end{deluxetable}

\begin{deluxetable}{lrcclcc}
\tablewidth{0pc}
\tablenum{4}
\tablecaption{Space Densities}
\label{table-lf}
\tablehead{
\colhead{Sp. Type} &
\colhead{N} &
\colhead{$\Phi$} &
\colhead{$\sigma_\Phi$} & 
\colhead{units} &
\colhead{$<\frac{V}{V_{max}}>$} &
\colhead{$\sigma_{<\frac{V}{V_{max}}>}$} 
}
\startdata
M8.0-M8.5  &   17. &    1.90 &    0.47 &                $10^{-3}$ stars pc$^{-3}$ &    0.56 &    0.07 \\ 
M9.0-M9.5  &   15. &    2.57 &    0.69 &                $10^{-3}$ stars pc$^{-3}$ &    0.71 &    0.07 \\ 
M8.0-M9.5  &   32. &    4.46 &    0.83 &                $10^{-3}$ stars pc$^{-3}$ &    0.63 &    0.05 \\ 
L0.0-L4.5  &    7. &    2.11 &    0.92 &                $10^{-3}$ stars pc$^{-3}$ &    0.53 &    0.11 \\ 
M8.0-L4.5  &   39. &    6.57 &    1.24 &                $10^{-3}$ stars pc$^{-3}$ &    0.61 &    0.05 \\ 
M8.0-M9.5  &   32. &    4.75 &    0.89 &     $10^{-3}$ stars pc$^{-3}$ mag$^{-1}$ &    0.63 &    0.05 \\ 
M8.0-L4.5  &   39. &    4.38 &    0.83 &     $10^{-3}$ stars pc$^{-3}$ mag$^{-1}$ &    0.61 &    0.05 \\ 
\enddata
\end{deluxetable}

\begin{deluxetable}{llrclc}
\tablewidth{0pc}
\tablenum{5}
\tablecaption{Flares}
\label{table-flares}
\tablehead{
\colhead{Name} &
\colhead{Sp.} &
\colhead{H$\alpha$} &
\colhead{Source} &
\colhead{Flare H$\alpha$} &
\colhead{Source}}
\startdata
2MASSW J0149089+295613 & M9.5 & 11.0 & L99 & 300 & L99 \\
2MASSW J2234138+235956 & M9.5 & 4.4 & KPNO & 20 & Keck \\
LHS 2397a & M8 & 15 & LCO & ?  & B91 \\
LHS 2243  & M8 & 1.3 & M94 & 44 & LCO \\
LP 475-855 & M7 & 7 & Keck &40.5 & LCO \\
\enddata
\tablecomments{Sources are:  LCO (Las Campanas, this paper),
KPNO (Kitt Peak, this paper), Keck (Keck, this paper), 
B91 (Bessell 1991), M94 (Martin, Rebolo, \& Maguzzu 1994), 
L99 (Liebert et al. 1999)
}
\end{deluxetable}

\end{document}